\documentclass[twocolumn, times, tighten]{aastex62}

\usepackage{graphicx}	
\usepackage{amsmath}	
\usepackage{amssymb}	

\usepackage{multirow}
\usepackage{longtable}
\usepackage{enumitem}

\usepackage[]{xcolor}

\received{2019 May 14}
\revised{2019 September 23}
\accepted{2019 September 24}

\shorttitle{Catalog of X-shaped Radio galaxies (XRGs)}
\shortauthors{Yang et al.}

\begin{document}

\title{Extended catalog of winged or X-shaped radio sources from the FIRST survey}

\correspondingauthor{Xiaolong Yang}
\email{yxl.astro@gmail.com}

\author[0000-0002-4439-5580]{Xiaolong Yang}
\affil{Kavli Institute for Astronomy and Astrophysics, 
Peking University, 
Beijing 100871, China}

\author{Ravi Joshi}
\affiliation{Kavli Institute for Astronomy and Astrophysics, Peking University, Beijing 100871, China}

\author{Gopal-Krishna}
\affiliation{Aryabhatta Research Institute of Observational Sciences (ARIES), Manora Peak, Nainital - 263002, India}

\author{Tao An}
\affiliation{Shanghai Astronomical Observatory, Key Laboratory of Radio Astronomy, Chinese Academy of Sciences, 200030 Shanghai, P.R. China}

\author{Luis C. Ho}
\affiliation{Kavli Institute for Astronomy and Astrophysics, Peking University, Beijing 100871, China}
\affiliation{Department of Astronomy, School of Physics, Peking University, Beijing 100871, China}

\author{Paul J. Wiita}
\affiliation{Department of Physics, The College of New Jersey, PO Box 7718, Ewing, NJ 08628-0718, USA}

\author{Xiang Liu}
\affiliation{Xinjiang Astronomical Observatory, Key Laboratory of Radio Astronomy, Chinese Academy of Sciences, 150 Science 1-Street, 830011 Urumqi, P.R. China}

\author{Jun Yang}
\affiliation{Department of Earth and Space Sciences, Chalmers University of Technology, Onsala Space Observatory, SE-439\,92 Onsala, Sweden}
\affiliation{Shanghai Astronomical Observatory, Key Laboratory of Radio Astronomy, Chinese Academy of Sciences, 200030 Shanghai, P.R. China}

\author{Ran Wang}
\affiliation{Kavli Institute for Astronomy and Astrophysics, Peking University, Beijing 100871, China}
\affiliation{Department of Astronomy, School of Physics, Peking University, Beijing 100871, China}

\author{Xuebing Wu}
\affiliation{Department of Astronomy, School of Physics, Peking University, Beijing 100871, China}
\affiliation{Kavli Institute for Astronomy and Astrophysics, Peking University, Beijing 100871, China}

\author{Xiaofeng Yang}
\affiliation{Xinjiang Astronomical Observatory, Key Laboratory of Radio Astronomy, Chinese Academy of Sciences, 150 Science 1-Street, 830011 Urumqi, P.R. China}



\begin{abstract} 
We present a catalog of 290 `winged' or X-shaped radio galaxies (XRGs) extracted from the latest (2014 Dec. 17) data release of the `Very Large Array Faint Images of the Radio Sky at Twenty centimeter' (VLA FIRST survey). We have combined 
these radio images with their counterparts in the TIFR GMRT sky survey (TGSS) at 150\,MHz, in an attempt to identify any low surface-brightness radio emission present in these sources. This has enabled us to assemble a sample of 106 `strong' XRG candidates and 184 `probable' XRG candidates whose XRG designation needs to be verified by further observations. The present sample of 290 XRG candidates is almost twice as large as the number of XRGs presently known. Twenty-five of our 290 XRG candidates (9 `strong' and 16 `probable') are identified as quasars. Double-peaked narrow emission lines are seen in the optical spectra of three of the XRG candidates (2 `strong' and 1 `probable'). Nearly 90\% of the sample is located in the FR\,II domain of the Owen-Ledlow diagram. A few of the strong XRG candidates have a rather flat radio spectrum (spectral index $\alpha$ 
flatter than $-0.3$) between 150\,MHz and 1.4\,GHz, or between 1.4\,GHz and 5\,GHz. Since this is not expected for lobe-dominated extragalactic radio sources (like nearly all known XRGs), these sources are particularly suited for follow-up radio imaging and near-simultaneous measurement of the radio spectrum. 
\end{abstract}

\keywords{galaxies: active --- galaxies: radio jets --- high energy astrophysics: supermassive black holes --- quasars: general --- radio continuum: galaxies: 
miscellaneous --- catalogs --- surveys}


\section{Introduction} \label{sec:intro}

Active galactic nuclei (AGNs) are long believed to be located at the centers of massive galaxies 
\citep[e.g.,][]{2013ARA&A..51..511K}. A plausible scenario to account for the enormous energy release 
from the AGN involves accretion of matter on to a super-massive black holes (SMBH) which, under 
suitable conditions, can also result in the launch of a pair of relativistic jets of non-thermal radio 
emission, which can extend up to mega-parsec dimensions \citep[e.g.,][]{2014ARA&A..52..589H, 
2017MNRAS.469.2886D}. Such galaxies, referred to as `radio galaxies' (RGs), often show a compact 
radio core flanked by a pair of `radio lobes'. Nearly always, powerful RGs are associated with 
elliptical galaxies \citep{2001A&A...375..791V}. Remarkably, a small minority of RGs is known to exhibit 
{\it two pairs} of fairly well collimated radio lobes, broadly forming an X-shaped morphology. These `winged' or 
`X-shaped' radio galaxies \citep[XRGs, e.g.,][]{1992ersf.meet..307L} form the topic of this work. 

XRGs constitute about 5 to 10\% of radio galaxies in the 3CRR catalog \citep{1984MNRAS.210..929L, 
1992ersf.meet..307L}. Based on an edge-darkened, or edge-brightened radio morphology of the brighter 
(i.e., primary) radio lobe pair, they are classified as Fanaroff-Riley type I (FR\,I) or type II (FR\,II) XRGs, 
respectively \citep{1974MNRAS.167P..31F}. Recall that whereas a pair of radio lobes, each having a `hot spot' 
near its outer edge, is seen in FR\,II sources, FR\,Is often exhibit a jet pair emanating from a prominent 
radio core, each jet forming an edge-darkened radio lobe. The primary lobe pair in a majority of XRGs 
in the \citet{1992ersf.meet..307L} sample belong to the FR\,II type, whereas the secondary lobes (`wings') in 
all known XRGs are of the FR I type. Also, interestingly, radio luminosities of XRGs are mostly 
near the FR\,I and FR\,II division \citep[$P_\mathrm{178MHz}\approx2\times10^{25}\,\mathrm{W\,Hz^{-1}\,
sr^{-1}}$, e.g.,][]{2009ApJS..181..548C, 2010MNRAS.408.1103L}. Their intermediate radio luminosities
may hint that XRGs represent a transitional morphology between the FR\,I and FR\,II types \citep{2010MNRAS.408.1103L}. 

The origin of XRGs is contentious and several models have been proposed to explain this phenomenon, 
as reviewed in \citet[][]{2012RAA....12..127G}. The three most discussed scenarios for their formation 
are: (1) diversion of the back-flowing synchrotron plasma of the radio lobes, upon impacting an 
asymmetric circum-galactic gaseous halo of the parent early-type galaxy; (2) precession of the (large-scale) 
twin-jets; and (3) spin-flip of the central SMBH. 

In the backflow model, the radio wings form due to diversion 
of the back-flowing lobe plasma whose subsequent outward expansion is aided by the buoyancy forces
exerted by a steep pressure gradient in the circum-galactic medium (CGM) of the parent galaxy \citep{1984MNRAS.210..929L, 
1995ApJ...449...93W, 2005ApJ...622..149K, 2009ApJ...695..755M}, the diversion occurring preferentially into pre-existing 
cavities/channels in the CGM \citep{1984MNRAS.210..929L, 2016A&A...595A..46M}. However, 
\citet{1992ersf.meet..307L} have cautioned that buoyancy driven backflow cannot propagate faster than 
the external sound speed and hence, the wings propelled by the buoyancy forces are not expected 
to be longer than the main radio lobes (which are known to advance supersonically). Since this is not always the 
case, a possible resolution 
may lie in the possibility that over-pressured cocoons can drive supersonic 
outflows of the backflowing lobe plasma, forming gigantic radio wings 
\citep{2002A&A...394...39C, 2011ApJ...733...58H, 2017A&A...606A..57R}. 
In order to realize the needed large over-pressure relative to the ambient medium, the jet's head 
would be required to first propagate out to sufficiently large distance from the nucleus. 

Observational support to the backflow deflection model comes from optical studies of the host galaxies 
of XRGs \citep{2002A&A...394...39C, 2009ApJ...695..156S, 2016A&A...587A..25G}. These studies have 
demonstrated that the radio axis defined by the two primary lobes displays a preference to align with the 
optical {\it major} axis of the host elliptical galaxy. The two wings (secondary lobes) show a strong tendency to align with the {\it minor} axis of the optical host, which is consistent with the expectation that the putative buoyancy driven expansion of the wings should occur along the maximum pressure gradient. In tune with this basic picture, XRGs are found to be associated predominantly with early-type galaxies whose ellipticity is abnormally high \citep[][and references therein]{2016A&A...587A..25G}. The strong statistical correlation observed between the host galaxy properties and the X-shaped radio morphology provides support to the over-pressured lobe scenario and this is corroborated by the available X-ray images of a few XRGs, which trace the (asymmetric) shape of their CGM \citep{2010ApJ...710.1205H}.

Taking a clue from the observed S-shaped radio morphologies of the nearby radio galaxies Centaurus A and 
3C\,272.1, \citet{1978Natur.276..588E} proposed that geodetic precession of the jet pair, conceivably due 
to a torque exerted by an external massive body, can explain the huge Z-symmetric radio trails extending 
from the extremities of the two lobes of the giant radio galaxy NGC\,326. The jet precession may also be 
caused due to a tilted and warped accretion disk \citep[see][and reference therein]{2006ApJ...653..112C}, 
or by a close passage of a neighboring galaxy \citep{1978MNRAS.185..527B, 2002MNRAS.330..609D}. 
Indeed, \citet{1980A&A....85..101B} reported that NGC\,326 is associated with a dumbbell galaxy, 
i.e., two nearly equally bright ellipticals within a common envelope, of which one is currently 
hosting jet activity and interacting gravitationally with its elliptical neighbor \citep{2001A&A...380..102M}. 
A study of $\sim$100 radio sources associated with dumbbell galaxies has in fact revealed markedly 
distorted radio structures in roughly a dozen of them \citep{1982AJ.....87..602W}. In this picture, 
the interacting neighbor may either be in a highly eccentric (or, unbound) orbit, or in a circular 
orbit. In the first case, impulsive gravitational interaction can lead to an inversion-symmetric 
distortion of the jet pair, resulting in an X-shaped or Z-shaped morphology. In the second case, the 
continued tidal interaction can cause periodic inversion-symmetric wiggles in the radio jet pair, giving 
rise to a helical radio morphology. A warped and tilted accretion disk can also be expected to form in 
SMBH binaries, leading to an X-shaped radio morphology \citep{1980Natur.287..307B}. One motivating factor 
behind this scenario is the discovery of double-peaked emission lines in the optical spectra of a 
few XRGs \citep{2007MNRAS.377.1215Z}. By modeling the radio morphological distortions observed in 3 XRGs 
(3C\,52, 3C\,223.1 and 4C\,12.03), \citet{2011ApJ...734L..32G} have estimated precession periods of the 
order of a million years; such a timescale would be consistent with the estimated dynamic ages of active 
radio lobes and the wings in XRGs \citep{2012A&A...544A..36M}.

In the `spin-flip' scenario, the wings of an XRG are regarded as the fossil synchrotron plasma of the 
earlier (pre spin-flip) lobe pair, while the observed misalignment of the currently `active' lobes from 
the axis defined by the wing pair is attributed to the jet's re-orientation, following a sudden tilt of the 
spin axis of the AGN's supermassive black hole (SMBH) resulting from coalescence of the two SMBHs 
belonging to the pair of merging galaxies \citep{2001PhDT.......173R, 2001A&A...377...23Z, 2002Sci...297.1310M}. 
However, it has also been argued that the wings could even form prior to the SMBH merger 
\citep{2002astro.ph.11503B, 2003ApJ...594L.103G, 2005MNRAS.364..583Z}. In a systematic study, 
\citet{2011A&A...527A..38M} have found that the host ellipticals of nearly half of the XRGs exhibit 
signatures of a starburst occurring around $10^9$ to $10^{9.5}$ years (1 to 3\,Gyr) ago. Hydrodynamical 
simulations of galaxy mergers suggest a time interval of about 2\,Gyr between a merger-induced starburst and 
the onset of AGN activity \citep{2008MNRAS.391.1137L}, in accord with the idea of a physical link between 
galaxy merger and XRG formation. There exist other observational evidences implicating galaxy mergers in 
the XRG formation. These include: (1) elliptical galaxies hosting XRG have SMBH of larger than average 
mass found using a control sample of elliptical galaxies \citep{2011A&A...527A..38M}; (2) Detection of 
X-ray cavities and a likely stellar shell in the XRG system 4C\,$+$00.58 \citep{2010ApJ...717L..37H}, as 
well as the detection of shells around the host galaxy of the XRG 3C\,403 \citep{2011MNRAS.410.1550R}. 
Also, a recent dynamical analysis of the hybrid radio structural distortion observed in the radio galaxy 
3C\,293 lends support to the rapid jet re-alignment scenario for XRGs \citep{2016A&A...595A..46M}. On the 
other hand, the merger-induced rapid reorientation model encounters difficulty in explaining the strong 
tendency for the wings of XRGs to align with the optical minor axis of host galaxy 
\citep[][and references therein]{2016A&A...587A..25G}.

It is important to bear in mind that none of the above models is capable of consistently explaining 
the entire gamut of observed properties of XRGs, as emphasized in a review of XRG models, which also
dwells upon a few alternative explanations \citep[][]{2012RAA....12..127G}. One of these alternatives 
invokes a collision between the radio jet pair with the (partial) shells around the host elliptical, that are 
believed to form in the process of galaxy merger (see \citealt{1983Natur.303..217G}; \citealt{1984A&A...141...61G, 2005MNRAS.364..583Z}). A possible substitute for the shells is the interstellar medium (ISM) of the massive host elliptical, which is set in rotation by a captured galaxy as it spirals inward in course of its merger with the massive elliptical \citep{2003ApJ...594L.103G}. 
A distinct merit of this model lies in its ability to provide a natural explanation for the Z-shaped morphology traced by the radio wings, a pattern highlighted in that study where it was  noted to be a fairly common feature of the XRGs with well-resolved radio maps. Yet another explanation proposed for the XRGs simply posits that the nucleus possesses not one, but two, SMBHs each of which ejects a jet pair grossly misaligned from the other (\citealt{2008MNRAS.390.1105L}, see also \citealt{2019arXiv190311632L}). This would have been the most direct explanation for XRGs, but for the difficulty it faces in explaining the observed Z-symmetry of the wings (see above), the observed preferential alignment of the wing pair with the optical minor axis of the host galaxy, as well as the fact that hot spots are never found in {\it both} pairs of radio lobes in 
XRGs \citep[see, e.g.,][]{2003ApJ...594L.103G}. Furthermore, VLBI observations of XRGs have so far provided no compelling evidence 
for a dual active nuclei inside the radio core \citep{2011MNRAS.410.2113B}.

The striking diversity of the different models proposed for the origin of XRGs makes them an extraordinarily 
interesting subclass of radio galaxies. In addition, they carry special astrophysical interest by being 
potential sources of gravitational waves, if indeed the `spin-flip' model is the correct description for at 
least a substantial fraction of XRGs \citep{1986ApJ...311..526H, 2001PhDT.......173R, 2002Sci...297.1310M, 2002astro.ph.11503B, 2003AIPC..686..201M, 2015ApJS..220....7R}, 
Thus, an improved estimation of the fraction of RGs that turn into XRGs via the spin-flip route would 
enable realistic predictions for the low-frequency gravitational wave background which is thought to 
pervade the universe \citep[e.g.,][and references therein]{2015ApJ...810L...6R}.

\begin{figure*}
\centering
\includegraphics[width=1\textwidth]{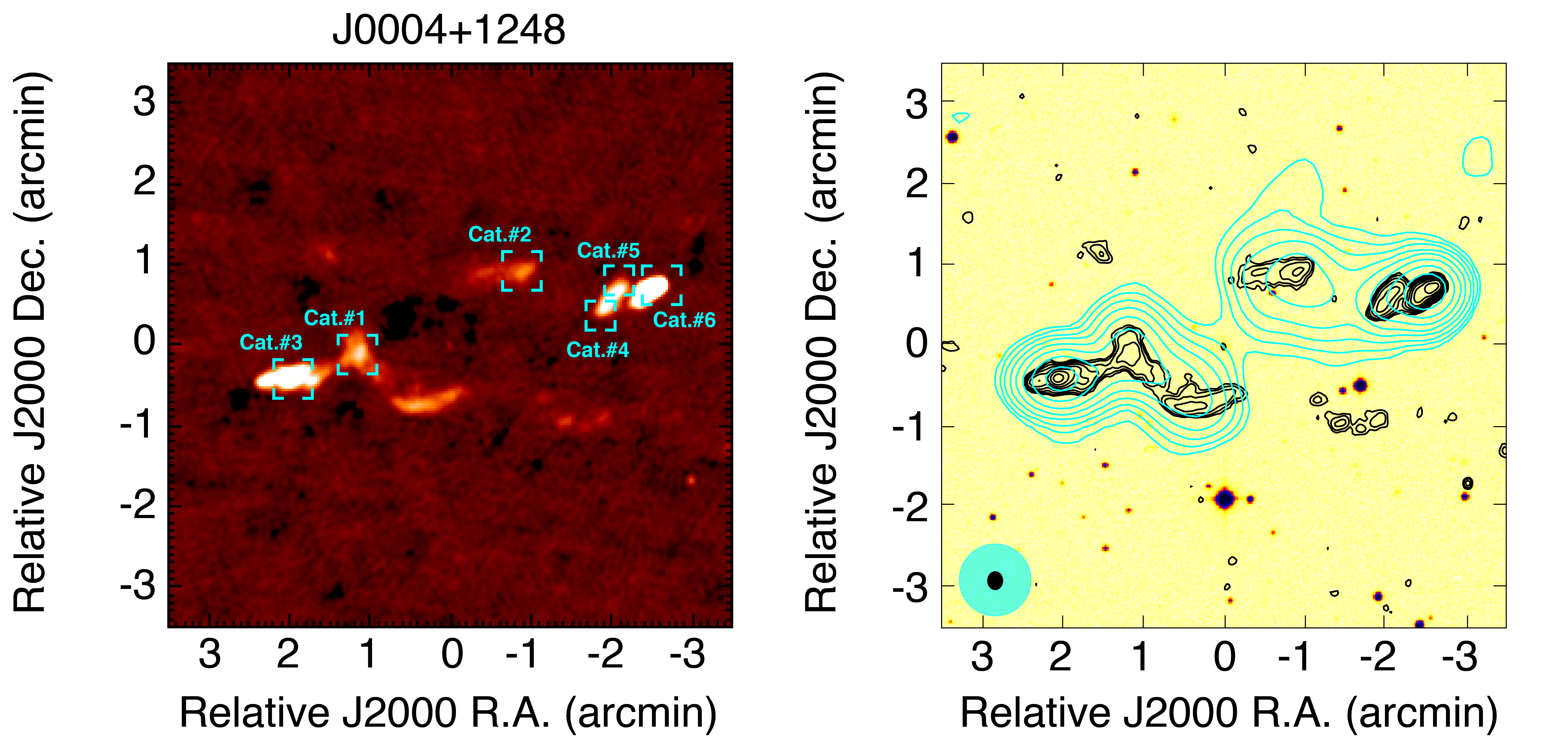}
\caption{Radio structure of a representative giant X-shaped radio source hosted by a galaxy at  the field center with
R.A.=00:04:50.27 (J2000), Dec.=$+$12:48:39.5 (J2000). Left Panel: the radiograph with all the FIRST catalog entries that satisfied our selection criteria marked (see Table \ref{tab:first}). 
Right panel: black contours are from the VLA FIRST survey at 1.4\,GHz, cyan contours are from the 
GMRT TGSS\_ADR1 survey at 150\,MHz. The TGSS\_ADR1 and FIRST contour levels are set at $0.6\times$ 
and $10\times$ (1, 2, 4, 8, 16, ...) mJy\,beam$^{-1}$, respectively. In the bottom-left corner, the synthesized beams (cyan for the 
TGSS\_ADR1 and black for the FIRST contours) are indicated by ellipses.}
\label{fig:eg}
\end{figure*}

The studies mentioned above are constrained due to the rather small sizes of the available XRG samples. 
The first major step towards rectifying this problem was taken by \citet[][]{1992ersf.meet..307L}. Later, \citet[][hereinafter C07]{2007AJ....133.2097C} identified 100 XRG candidates from a systematic search in the FIRST survey catalog \citep{1995ApJ...450..559B}. Their search was limited to sources with radio major axis size larger than 15\,arcsec, expectedly resulting in a significant under-representation of the XRG population in their list. Furthermore, the available optical and radio details are highly incomplete even for that sample. For instance, out of the 100 XRGs in the C07 sample, only $\sim$50\% have been observed spectroscopically (\citealt{2009ApJS..181..548C}, hereinafter C09, \citealt{2012A&A...544A..36M}) and only 53 out of the 100 XRGs have been taken up for investigating the host galaxy properties \citep{2016A&A...587A..25G}. 
Similarly, analysis of the SMBH mass and starburst history \citep{2011A&A...527A..38M} have been reported for just 29 out of the 100 XRGs, and merely 12 of them are covered in the follow-up study by \citet{2012A&A...544A..36M}. Therefore, in order to enlarge the scope of the studies of XRGs, it is desirable to extend the XRG search campaign initiated by \citet{2007AJ....133.2097C}. We have undertaken such a search program by lowering the radio source size threshold for XRG candidates from 15$^{\prime\prime}$ employed in that study, to 10$^{\prime\prime}$.
Based on this revised size threshold, we present here a new catalog of candidate X-shaped 
radio sources, drawn from the latest data release of the FIRST survey at 1.4\,GHz \citep{1995ApJ...450..559B}. 
In Section \ref{sec:cat}, we briefly comment on the existing XRG catalogs. Details of our selection 
procedure and the main results of our campaign are presented in Section \ref{sec:first}. The host galaxy 
identification and radio properties are described in Sections \ref{sec:host} \& \ref{sec:radio}, 
followed by a brief discussion (Section \ref{sec:dis}) and a summary of the present results (Section \ref{sec:sum}). Throughout this paper, we assume a $\Lambda$-$ \mathrm{CDM}$ cosmology with parameters $H_0=70\,\mathrm{km\,s^{-1}\,Mpc^{-1}}$, $\Omega_\mathrm{\Lambda}=0.7$ and $\Omega_\mathrm{m}=0.3$.

\section{Existing lists and definition of the `winged' or `X-shaped radio sources} \label{sec:cat}

As a class, X-shaped extragalactic radio sources were first discussed by \citet{1992ersf.meet..307L}, based on their list of 11 XRGs. Of these, the primary lobe pair showing a clear FR\,II radio morphology with well-defined hot spots is seen in 7 sources. \citet{2007AJ....133.2097C} collected another 8 XRGs through a literature search, raising the sample to 19 XRGs. The first systematic search for the XRGs was undertaken by \citet[][C07]{2007AJ....133.2097C}, based on the VLA FIRST survey \citep{1995ApJ...450..559B}. He compiled an initial list of 100 XRG candidates showing a `winged' or X-shaped radio morphology, by visually inspecting 1648 sources in which at least a hint of inversion symmetric radio lobe structure was present and the primary lobes extend more than 15 arcsec. Recently, \citet{2018ApJ...852...47R} have reported JVLA multi-array (mainly A-array) radio continuum imaging in the L- and/or C-band of XRG candidates from C07. Based on these observations, \citet[][]{2018ApJ...852...48S} have subsequently identified 12 sources as  S- or Z-shaped radio galaxies. Remarkably, more than 75\% of the C07 sample has turned out to be bona-fide XRGs and hence very useful for probing the question of origin of XRGs \citep[see][]{2009ApJS..181..548C, 2011A&A...527A..38M, 2012A&A...544A..36M, 2016A&A...587A..25G}. 

A more recent XRG search is reported in \citet{2011ApJS..194...31P} who applied an automated morphological classification scheme to the FIRST radio sources. Adopting a separation cut of 0.96\,arcmin, they classified the radio sources into singles, doubles, triples, and groups of higher membership count. They also visually inspected 7106 higher count group members, thus finding 156 candidates for X-shaped radio morphology. Out of these, 21 sources had already been reported in C07 and one object, 3C\,315, is a well-known XRG (see below), leaving 134 new XRG candidates (this is $\sim$2\% of the sources classified by them as higher-count groups). 
Two of us (RJ and XLY) have visually inspected the 156 XRG candidates reported in \citet{2011ApJS..194...31P}, in search of an unambiguous X-shaped morphology with well defined wings (see, Section \ref{sec:class} for details of our selection procedure) and assessed 43 of them to be strong XRG candidates, and the remaining systems as the probable candidates. Out of these 43 strong candidates, 18 sources are common to C07, and FCG J151340.0$+$260730 (3C\,315) is the archetypal, X-shaped radio galaxy known for over 4 decades \citep{1969MNRAS.145...31M, 1974A&A....34..341H}. Thirteen of the remaining 24 sources were also picked in the present search for strong XRG candidates in the 2014 Dec. 17 release of the FIRST catalog and these are marked with an asterisk in Column 1 of Table \ref{tab:strong}, while 11 were not because they did not fulfill our criterion that the radio major and minor axes should be greater than 10 and 5 arcsec, respectively - see section \ref{sec:basic}. 

\begin{table}
\renewcommand\arraystretch{0.8}                                                                                                                                                                              
\centering                                                                                                                                                                                 
\caption{The FIRST catalog entry for each component in the field J000450.27+124839.5 (see Figure \ref{fig:eg}).} \label{tab:first}                                                                                                                                                                                                                                                                                                                                                   

\begin{tabular}{cccc}

\toprule
Component &  Peak Flux  &  Major Axis$^a$  &  Minor Axis$^a$    \\
        &  (mJy/beam) &  (arcsec)      &  (arcsec)        \\
\hline
Cat.\#1  &  7.80       &  22.88         &  15.44           \\
Cat.\#2  &   5.72       &  14.47         &  7.64          \\
Cat.\#3  &   20.29      &  23.41         &  10.98           \\
Cat.\#4  &  12.56      &  12.03         &  7.19            \\
Cat.\#5  &  12.60      &  14.25         &  7.59            \\
Cat.\#6  &  56.57      &  11.28         &  6.79            \\
\hline
\end{tabular}

\textbf{Notes}: $^a$ The deconvolved synthesized beam major and minor axes.                                                                                                                                                                          
\end{table}                                                                                                                                                                      

It is worth emphasizing that even though the visual inspection approach for picking XRG candidates is 
beset with subjectivity, it does yield promising XRG candidates, as validated by the high ($>$75\%) confirmation 
rate of the XRG candidates reported in C07 \citep[e.g.,][]{2018ApJ...852...47R, 
2018ApJ...852...48S}. Recalling the standard definition, the XRGs are a  subset of radio sources which 
exhibit an additional (fainter) pair of radio lobes displaying an inversion-symmetric 
configuration. These `secondary' radio lobes, often called `wings', are aligned at a fairly large 
angle from the main radio axis defined by the primary lobe pair. Conventionally, the readily accepted 
examples of XRGs are those in which the wings extend to at least 80\% of the size of the primary lobes. 
But, since the wings always lack a brightness peak (hot spot) near the extremity, their measured radio 
extents (and hence the robustness of their XRG classification) would depend sensitively on the depth 
and spatial resolution of the radio map. For instance, although only one wing of the well known XRG 
3C\,63 is apparent in the 1.4 GHz VLA B-array map \citep{1988ApJS...68..643B}, both wings clearly 
stand out in its subsequent VLA A-array image at 1.4 GHz \citep{1998ApJS..119...25H}. Similarly, the 
large, prominent wings present in the radio maps of the well known XRGs 3C\,192 and 3C\,379.1 
\citep{1985ApJ...291...52M, 1988ApJS...68..643B, 1999MNRAS.304..271D} only appear as short 
extensions in their earlier shallower radio maps \citep{1997MNRAS.291...20L, 1985ApJ...291...52M}. 
Furthermore, it now appears quite plausible that XRGs are morphological cousins of Z-shaped radio 
galaxies, in which the oppositely directed wings launch out from the main radio axis at locations that are 
relatively close to the host galaxy \citep[e.g.][]{2003ApJ...594L.103G}. Thus, in isolation, the two wings 
can often be described as a Z-shaped double radio source \citep[][]{2003ApJ...594L.103G}. Guided by 
all these considerations and in order to minimize missing out genuine XRGs, we have resorted to a 
somewhat less conservative approach in assembling the present catalog of XRG candidates, by also 
including the sources showing short wings (or even a one-sided wing), or just a hint of X-shaped radio 
structure. This was followed up by classifying them as `strong' and `probable' XRG candidates (section 
\ref{sec:class}). Clearly, the latter would inevitably need higher quality radio images for validation as 
genuine XRGs.

\begin{figure*}
\centering
\includegraphics[width=1\textwidth]{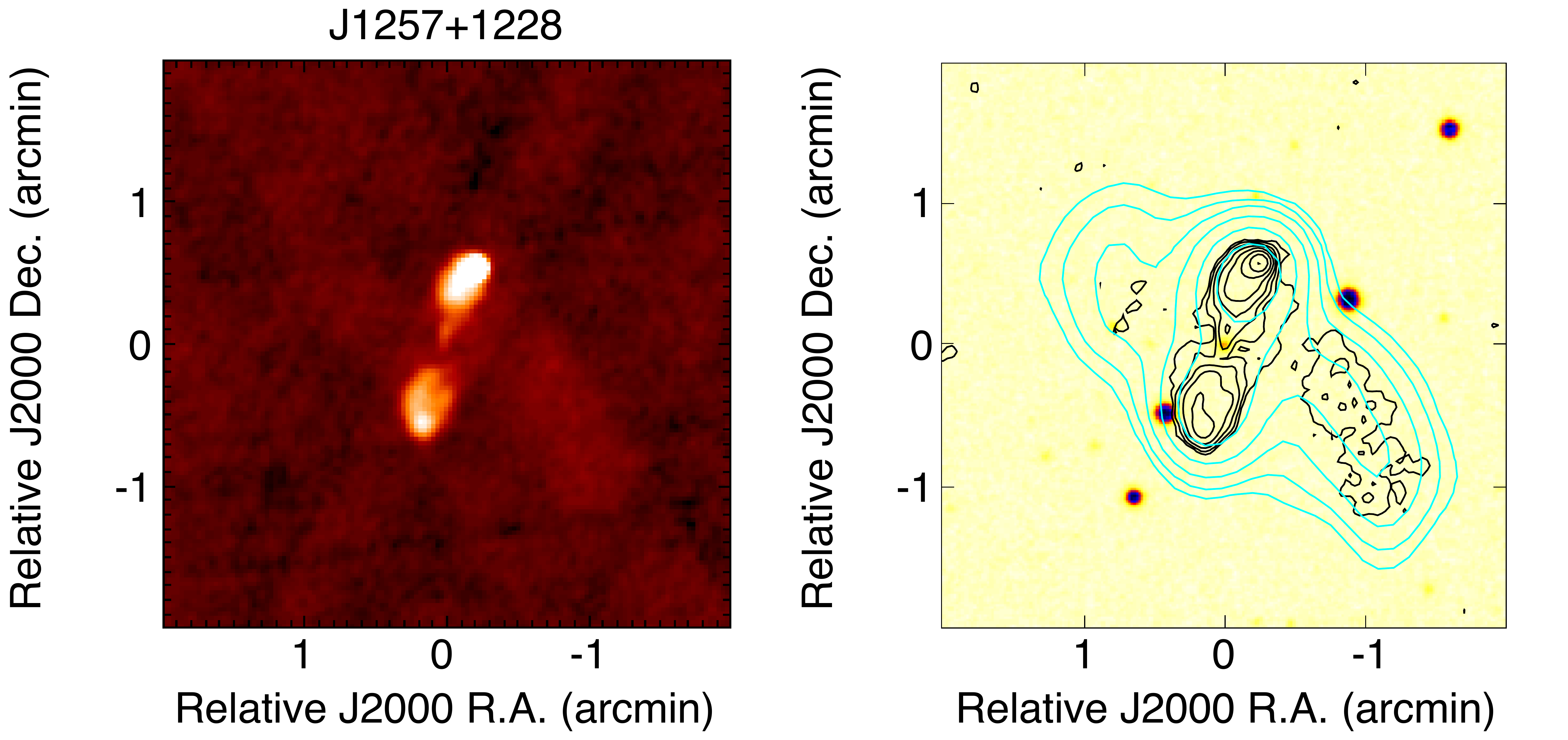}
\caption{An example of a `strong' XRG candidate in the present catalog: J1257$+$1228. 
The left panel shows the FIRST grayscale image and the right panel shows the FIRST (black) and TGSS\_ADR1 contours (cyan) overlaid on the DSS red filter image.
The contours are plotted at (1, 2, 4, 8, 16, 32, ...)$\times3\sigma$. The rms ($1\sigma$) noise value is 0.14 \,mJy\,beam$^{-1}$ for the FIRST contours, and 3.5\,mJy\,beam$^{-1}$ 
for the TGSS\_ADR1 contours. This example underscores the importance of combining the 1.4\,GHz maps with their 150\,MHz counterparts.}
\label{fig:tgss}
\end{figure*}

\section{Searching for X-shaped radio sources in the latest release of the {\it FIRST} survey} \label{sec:first}

We have assembled the present catalog of 290 XRG candidates by searching in the latest VLA FIRST survey data release version 14Dec17 \footnote{Data released on 2014 Dec. 17}. 
This version covers 10,575 square degrees of the sky (8,444 square degrees in the northern and 2,131 square degrees in the southern hemisphere) and contains 946,432 radio sources, including those reported in the earlier data releases from 1993 through 2011. The survey achieved a typical rms noise of 0.15 mJy and a resolution of $\sim 5^{\prime\prime}$ at 1.4 GHz \citep{1995ApJ...450..559B}. The main product of the FIRST survey is a radio intensity map of the sky, using which a catalog of discrete radio sources has been built from the coadded images \citep{1997ApJ...475..479W}. The catalog also lists for each entry the peak and integrated flux densities and angular size at 1.4 GHz, derived from fitting two-dimensional Gaussians. Note that in the FIRST catalog, an extended radio source can often have multiple entries \citep[][see also Figure \ref{fig:eg}]{2003JEI....12..398P, 2006ApJS..165...95P}. As estimated by \citet{2011ApJS..194...31P}, $\sim$30\% of the entries are actually individual multi-component radio sources, which offers a huge data base for morphological studies.

\subsection{Basic sample selection} \label{sec:basic}

Given the enormous size of the FIRST catalog and the abundance of multi-component systems in it (see, e.g., Figure \ref{fig:eg} and Table \ref{tab:first}), examining 
the map of each source is not a practical option. Therefore, we first extracted a subset of XRG candidates 
by demanding (1) a peak flux density sufficiently high to realize an acceptable dynamic range, and (2) a 
radio extent sufficiently large to reveal the basic morphological features. Below, we describe our sample 
selection procedure in some detail.

The characteristic feature of XRGs is two (mis-aligned) pairs of radio lobes/jets, such that there is a large angular 
offset between the axes defined by the brighter (primary) lobe pair and by the two secondary lobes (i.e., 
`wings') whose surface brightness is usually much lower. Only the primary lobes are known to exhibit an 
edge-brightened morphology, i.e., a hot spot typically situated near the lobe's extremity. Considering the 
typical rms noise of the FIRST maps (0.15\,mJy\,beam$^{-1}$), a 3$\sigma$ detection threshold would correspond 
to a minimum peak flux density of 0.45\,mJy\,beam$^{-1}$. Here, we make a reasonable assumption that the peak 
brightness of the hot spots in the primary lobes is about 10 times the (average) surface brightness of the 
wings. Accordingly, as the first selection filter, we set a lower limit of 5\,mJy\,beam$^{-1}$ for the peak 
flux density at 1.4\,GHz. This would normally permit a minimum dynamic range of $\sim$33:1 and one may thus 
reasonably expect to detect any associated wings (see section \ref{sec:host}). Next, in order to minimize 
missing out any wings we need to ensure that a given source is fairly well resolved with the $\sim$4.3 arcsec 
synthesized beam (VLA B-array) of the FIRST survey. Hence our second selection filter is that the radio major ($\theta_{maj}$) 
and minor ($\theta_{min}$) axes of the fitted Gaussian should be larger than 10$^{\prime\prime}$ and 
5$^{\prime\prime}$, respectively. In this way, the present compilation is an extension of the XRG search 
reported in C07 (see their section 3.1). Application of the above two selection filters left us with a basic list 
of 5128 sources. Of these, 2350 sources have $10^{\prime\prime} < \theta_{maj} \leqslant 13^{\prime\prime}$, 935 sources have $13^{\prime\prime} < \theta_{maj} \leqslant 15^{\prime\prime}$ and the remaining 1843 sources have $\theta_{maj}>15^{\prime\prime}$.

\subsection{Visual inspection of the radio maps of the short-listed sources}

In the next step, we queried the radio field for each of the 5128 short-listed sources, from the online archive of the FIRST survey\footnote{http://sundog.stsci.edu/index.html} \citep{1995ApJ...450..559B}, with an image size of 6$\times$6\,arcmin$^2$. Given the usually complex (multi-component) radio structures of known XRGs, we decided to also make use of their 150\,MHz radio continuum images reported in the First alternative data release of TIFR GMRT Sky Survey\footnote{https://vo.astron.nl/tgssadr/q\_fits/imgs/form} \citep[TGSS\_ADR1][]{2017A&A...598A..78I}. Its combination of angular resolution ($\sim$ 25\,arcsec) and a low frequency (150\,MHz) is better suited for picking up diffuse emission. 
The source J1257$+$1228 is a good example here (see Figure \ref{fig:tgss}). It is impossible to tell if it is an X-shaped winged source based on visual inspection of the FIRST grayscale image alone, whereas according to the FIRST-based contour map, J1257$+$1228 is `half X-shaped'. Only the TGSS contour map confirms it is in fact an X-shaped source. Radio contour maps and pseudo-color images were then generated to facilitate the visual inspection (see Appendix Figure \ref{fig:xrgs} for the `strong' XRG candidates). For each field, we measured the rms noise and used it for setting the base level for the contour plotting. We initially used $3\sigma$ level and slightly fine-tuned it in individual cases, so as to minimize the confusion arising from side lobes in the map. For the color images, we have carefully set the brightness to emphasize any wing-like features.

The above procedure was then followed up with two of the co-authors independently inspecting the radio structure of each short-listed source and their mutual concurrence was treated as the trigger for admitting a given source as a preliminary XRG candidate. In the final round, radio images of all such sources were individually inspected by three of the co-authors, which led to the final list of 290 XRG candidates.
Depending on the level of consensus among the 3 co-authors, these XRG candidates were placed in `strong' (106) or `probable' (184) categories (Table \ref{tab:strong} and Table \ref{tab:prob}). The classification is further discussed in section \ref{sec:class}. It may be noted that only 25 of these XRG candidates appear in the list of \citet{2011ApJS..194...31P}. Note also that since the present XRG sample is meant to be an extension of the C07 XRG sample, we have decided to retain these 25 XRG candidates, of which 13 systems belong to our list of strong XRG candidates. Those sources are marked with an asterisk in the first column in Table \ref{tab:strong} and Table \ref{tab:prob}.

\section{Identification of the host galaxy and its properties} \label{sec:host}

In this key step, we first defined for each source a likely position of the active core, near the symmetry 
centre of its radio structure. The radio contours were then overlaid on the Sloan Digital 
Sky Survey (SDSS) SDSS $i$-band and/or Digital Sky Survey (DSS) red-filter image of the source. 
For most ($\sim$80\%) of our XRG candidates, an optical counterpart could thus be found close to 
the expected location. For the remaining sources, we have simply used the estimated radio symmetry 
center as the coordinates of the (undetected) optical counterpart. These coordinates for our 106 `strong' and 184 
`probable' XRG candidates are listed in columns 2 and 3 of Table \ref{tab:strong} and Table 
\ref{tab:prob}, respectively. In Figure \ref{fig:xrgs} (see Appendix) we have displayed the 
FIRST 1.4\,GHz images of our 106 `strong' XRG candidates, the corresponding TGSS\_ADR1 
150\,MHz image is shown in the right panel of each subplot, except for J0710$+$3546 which is 
not covered in that survey. The centre of each image coincides with the above estimated most plausible position of the host galaxy. 
Interestingly, a radio core is detected in only $\sim10$ of our strong XRGs. 

\begin{figure}
\centering
\includegraphics[width=0.48\textwidth]{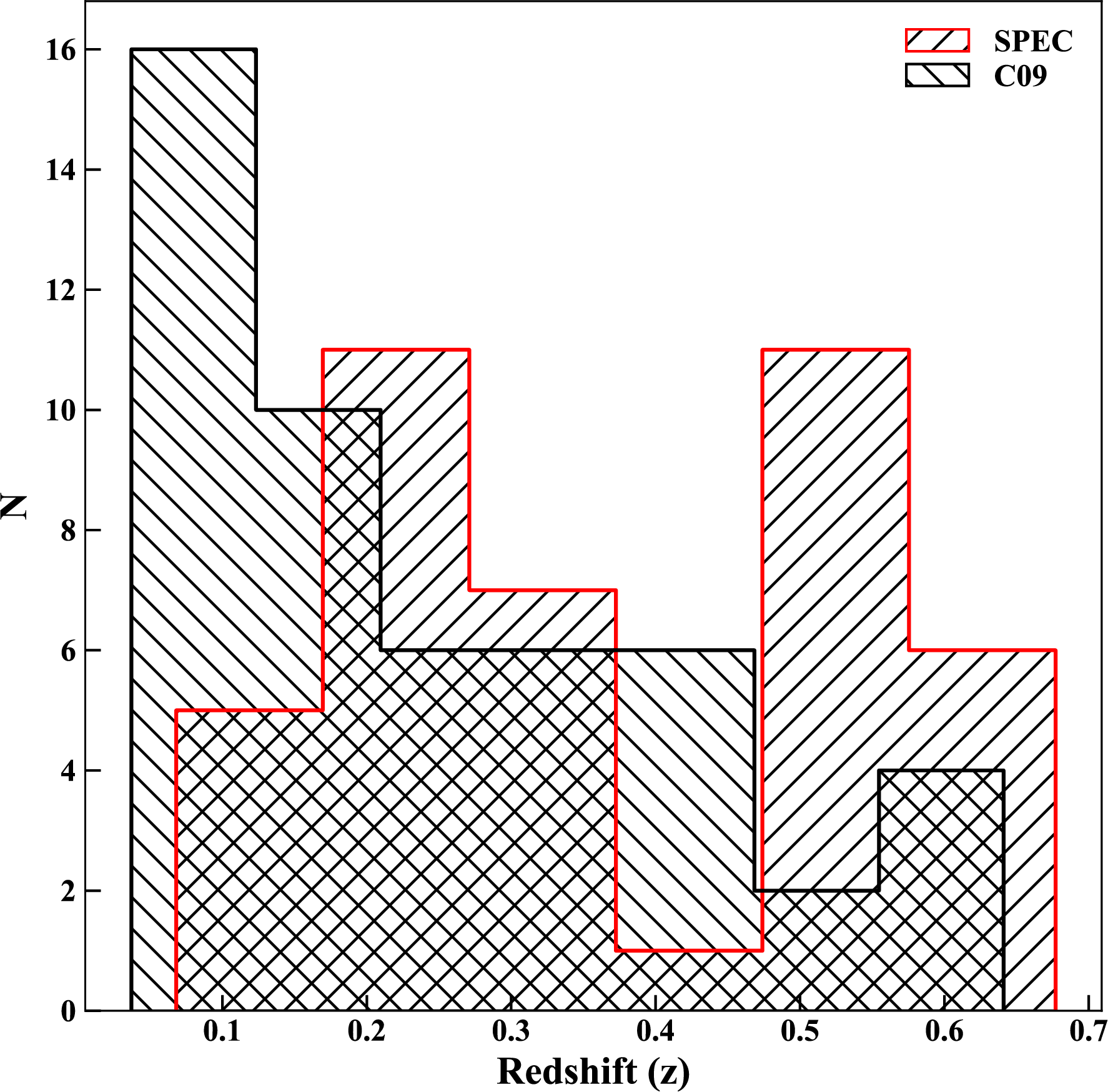}
\caption{Distributions of spectroscopically measured redshifts available for the 40 strong candidates in the present XRG catalog
(SPEC, red-lined histogram) and for the
50 XRGs from C09 (black-lined histogram).}
\label{fig:zhist}
\end{figure}

Spectroscopic or, alternatively, photometric redshifts were taken from the NED and/or
SDSS databases. Spectroscopic redshifts could be found for 41 ($\sim$39\%) of our 106 `strong' 
XRG candidates and 61 ($\sim$33\%) of our 184 `probable' XRG candidates. For the remaining `strong' 
candidates, photometric redshifts were taken from the SDSS archive; these are based on 
the SDSS photo-tree method. In Table \ref{tab:strong} and Table \ref{tab:prob}
we list the redshift values (Column 4) marked as `SPEC' and `PHOT ' for the spectroscopic and photometric redshifts (Column 5),
SDSS $r$-band apparent magnitude (Column 6) and the corresponding absolute magnitude 
(Column 7) for our 106 `strong' and 184 `probable' XRG candidates,  respectively. 
Figure \ref{fig:zhist} compares the histograms of spectroscopic redshifts for our strong XRG candidates
and for the 50 XRGs from \citet{2009ApJS..181..548C}. It is evident that our strong XRGs are systematically more
distant and span a larger range in redshift (from $\sim0.06$ to $\sim0.7$, with a median redshift $z\sim0.37$), 
as compared to the 50 XRGs in C09, for which the median redshift is $z\sim0.25$. 
This difference is not unexpected, given that our selection 
procedure admits sources of smaller radio (angular) extent (section \ref{sec:first}).

The $R$-band absolute magnitudes ($\mathrm{M}_R$) of our XRG candidates were computed from the SDSS 
$r$-band apparent magnitudes, by applying the $k$-correction from  \citet{2007AJ....133..734B}. The 
average $\mathrm{M}_R$ of our 41 spectroscopically identified `strong' XRG candidates is found to be $-$22.3 with an rms scatter of 0.8. It is interesting that these XRG candidates with spectroscopic redshifts contain nine quasars (marked as `SPEC/Q' in column 5 of Table \ref{tab:strong} ). It may be 
mentioned that X-shaped quasars are known to be quite rare; the first one was reported by 
\citet{2003AJ....126..113W} (4C$+$01.30 at $z$ = 0.132). Subsequently, one more X-shaped quasar, WGA J2347$+$0852, was reported by \citet{2006ApJ...637..183L}. In C07, 4 quasars were identified among their 36 spectroscopically identified XRG candidates. Recently, \citet{2018ApJ...852...48S} have confirmed a total of 12 quasars in the C07 sample. It may also be noted that just one X-shaped quasar (4C$+$01.30) is reported to exhibit a double-peaked (broad) emission line system \citep{2007MNRAS.377.1215Z}. The present catalog of XRG candidates contains another three such rare objects showing double-peaked narrow emission lines: the quasars J0818$+$1508 and J1554$+$3811 belong to the `strong'  XRGs (Table \ref{tab:strong}), while  J1247$+$1948 is a `probable' XRG candidate (see Table \ref{tab:prob}). Since the presence of double-peaked emission lines in X-shaped radio sources may be taken as a clue favoring the binary/dual black-hole model \citep[e.g.][]{2008MNRAS.390.1105L, 2019arXiv190311632L}, such rare objects are good candidates for confirming a SMBH pair/binary, via high-resolution VLBI. A more detailed analysis of the radio-optical properties of the present XRG candidates is underway and will be presented elsewhere (Joshi et al., in prep.).

\begin{figure}
\centering
\includegraphics[width=0.48\textwidth]{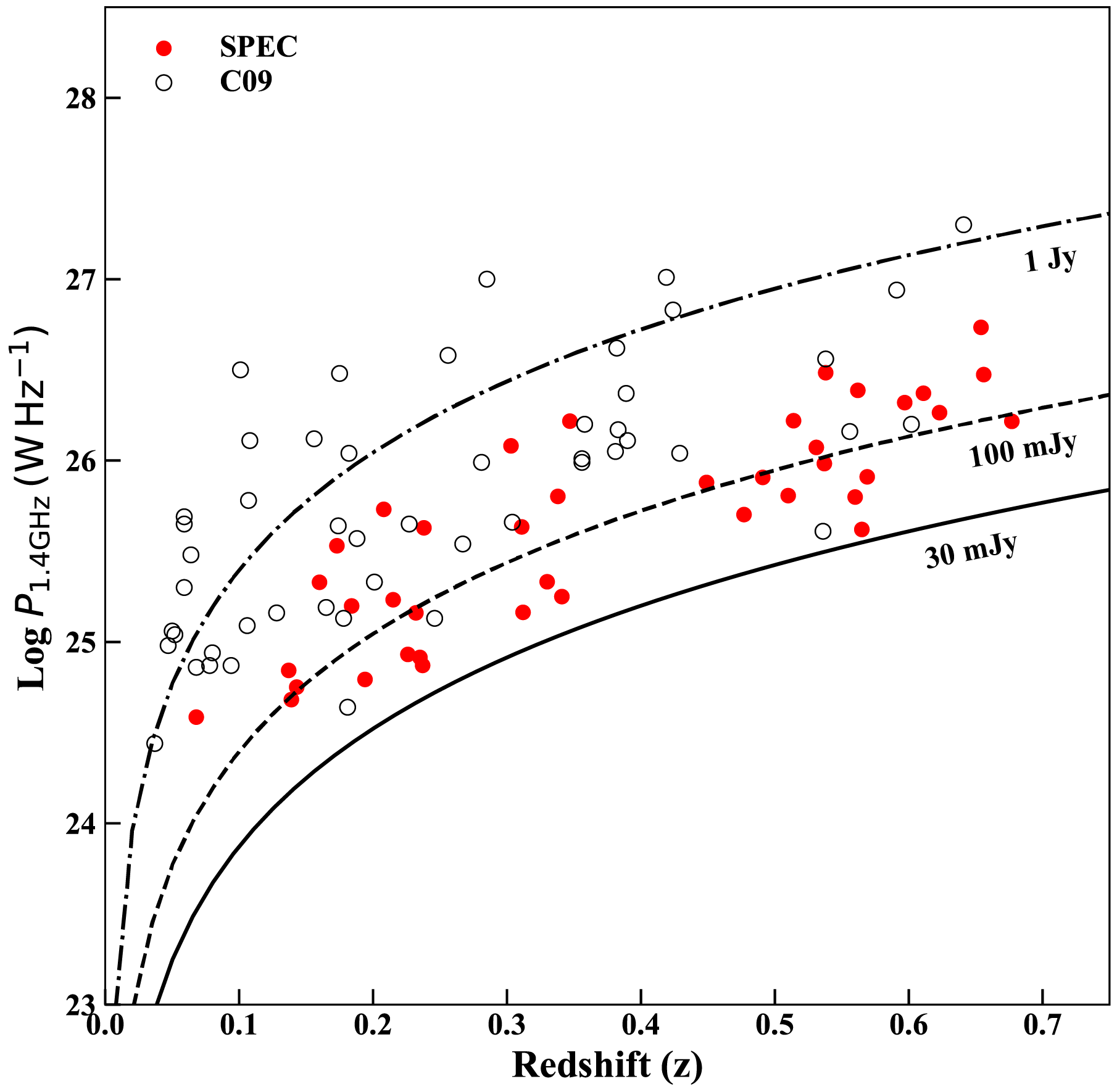}
\caption{Radio luminosity - redshift diagram for the 40 XRG candidates with available spectroscopic redshift and 1.4\,GHz radio luminosity (SPEC). 
The 50 XRG candidates with available spectroscopic redshifts \citep[][]{2009ApJS..181..548C} are shown with black 
open circles, while our 40 strong XRG candidates are shown with red filled circles. The solid, dashed and dot-dashed 
curves correspond,  respectively, to flux densities of 30\,mJy, 100\,mJy and 1\,Jy, at 1.4 GHz.}
\label{fig:lz}
\end{figure}

\section{Radio properties of the present catalog of XRG candidates} \label{sec:radio}
For the present XRG sample, we have gathered the radio flux density information at several 
frequencies: at 150\,MHz from the TGSS\_ADR1 catalog \citep{2017A&A...598A..78I}, at 1.4\,GHz 
from the NRAO VLA Sky Survey \citep[NVSS,][]{1998AJ....115.1693C} and at 4.85\,GHz/5\,GHz 
from the Parkes-MIT-NRAO surveys \citep[PMN,][]{1995ApJS...97..347G}, the MIT-Green Bank 
5\,GHz Survey \citep[MIT-GB,][]{1986ApJS...61....1B} and from the Green Bank 4.85\,GHz 
survey \citep{1991ApJS...75....1B}. These flux densities are listed in columns 8 - 10 of Table \ref{tab:strong} 
and Table \ref{tab:prob} for our `strong' and `probable' XRG candidates, respectively. We also 
list the 1.4\,GHz radio luminosity for all the XRG candidates, based on spectroscopic, or, 
alternatively, photometric redshifts. In calculating radio luminosity, we assumed a spectral index $\alpha=-0.8$ (see below). The average luminosity of our strong XRG candidates with 
available spectroscopic redshifts is found to be $log(P_\mathrm{1.4GHz})=25.73\,\mathrm{W\,Hz^{-1}}$, 
with an rms uncertainty of 0.65. This is in agreement with the value given in C07 and is close to the 
division between the FR\,I and FR\,II types \citep{1974MNRAS.167P..31F}. Figure \ref{fig:lz} 
displays the 1.4\,GHz luminosity vs redshift for our strong XRG 
candidates with known spectroscopic redshifts. It is seen that our XRG candidates have 
systematically lower radio luminosity compared to the previously reported XRGs \citep{2009ApJS..181..548C}. 
Interestingly, among our `strong' XRG candidates four objects, namely, J0004$+$1248, J0028$-$0026, J0932$+$1611 and J1324$+$4334, have a physical size of about $0.80, 0.90, 0.83,0.91$\,Mpc, respectively,  in the FIRST 1.4\,GHz map (measured as the separation 
between the ends of the two active radio lobes) and these belong to the family of giant radio galaxies \citep{1974Natur.250..625W, 2019arXiv190400409D}. Figure \ref{fig:eg} shows one of these 4 giant X-shaped radio galaxies, J0004$+$1248.

\begin{figure}
\centering
\includegraphics[width=0.48\textwidth]{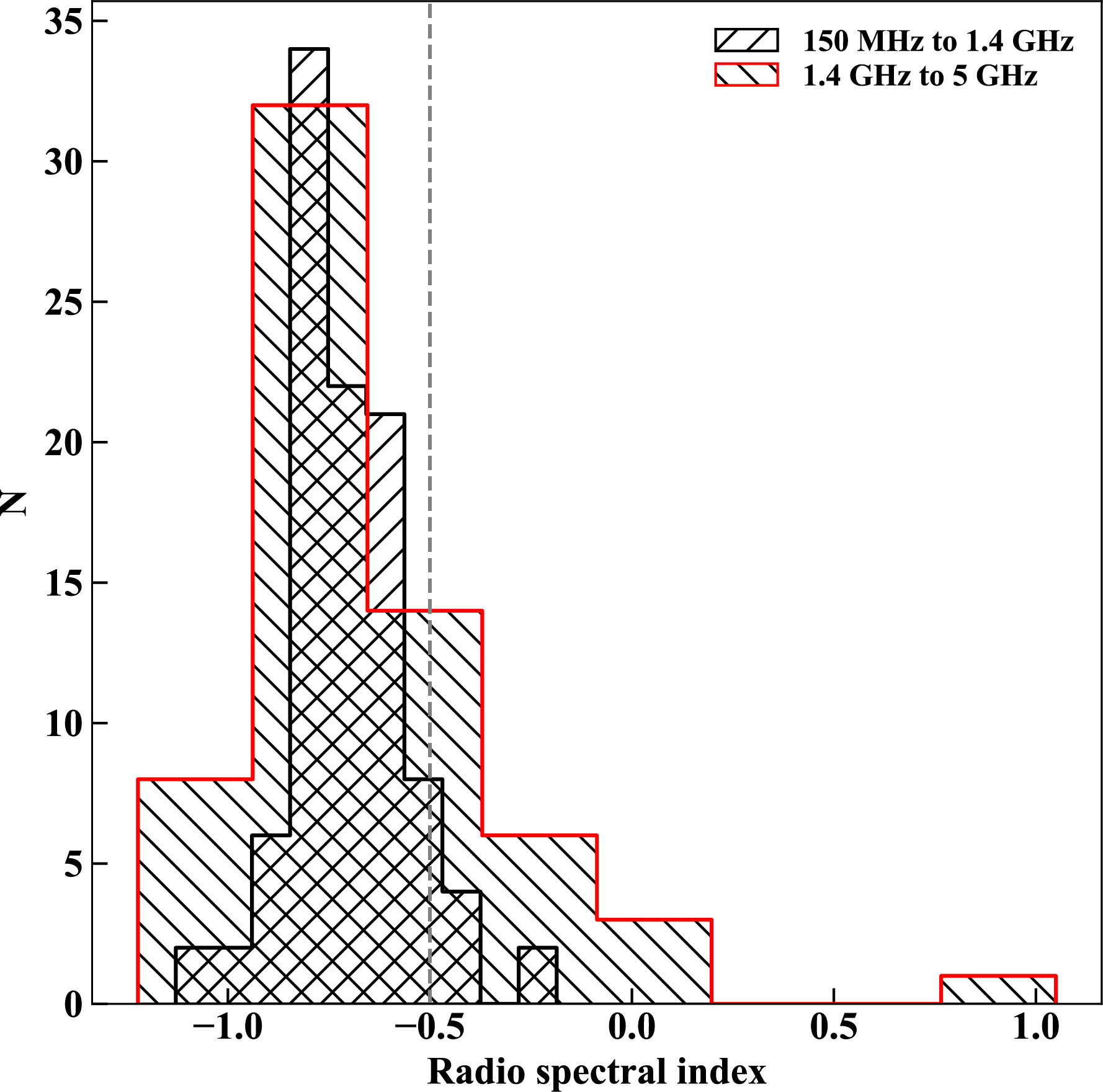}
\caption{Spectral index distributions for our strong XRG candidates. The histogram in red color refers to the spectral index (150\,MHz -- 1.4\,GHz ) values available  for 101 strong XRG candidates.The histogram in black color refers to the spectral index (1.4 GHz -- 5 GHz) values available  for 64 of our  strong XRG candidates.
The vertical dashed line marks the division between the steep (left side) and flat spectra (right side).}
\label{fig:si}
\end{figure}

For our strong XRG candidates, we have determined the spectral index ($S_i \propto \nu^{\alpha}$) 
between 150\,MHz and 1.4\,GHz ($\alpha_{0.15}^{1.4}$), and also between 1.4\,GHz and 5\,GHz ($\alpha_{1.4}^{5}$), based on the flux densities listed in Table \ref{tab:strong}. Thus, $\alpha_{0.15}^{1.4}$ could be determined for 101 of our strong XRG candidates, out of which 9 ($\sim$9\% ) sources are found to have a flat radio spectrum ($\alpha_{0.15}^{1.4}>-0.5$). In addition, $\alpha_{1.4}^{5}$ could be measured for 64 of our strong candidates, with 14 ($\sim$22\%) of them showing a flat radio spectrum and 3 out of these 14 actually showing an inverted radio spectrum (i.e. $\alpha_{1.4}^{5} > 0$). The XRG candidates showing a flat/inverted radio spectrum would be particularly valuable for probing the dual-AGN scenario for the origin of X-shaped morphology \citep[e.g.,][]{2007MNRAS.374.1085L}.  Histograms of the two spectral indices, for our strong XRG candidates are shown in Figure \ref{fig:si}. Both distributions peak near $\alpha = -0.8$.

\section{Discussion}  \label{sec:dis}

\subsection{Classification as `strong' or `probable' XRG candidates} \label{sec:class}

The vast majority in the `strong' 
category is likely to be confirmed as bona-fide XRGs, with a pair of radio wings comparable in 
extent to their active counterparts, the two primary radio lobes. 
In this section we briefly comment on the justification for including some sources in the present 
catalog, despite only a marginal evidence for a secondary lobe pair (wings) in their existing
radio maps. This stems from the recognition that clear visibility of wings can be hampered due to 
a number of factors related to the source evolution/orientation and their directional offset from the 
primary lobe pair. These difficulties may be compounded by observational limitations related to  
angular resolution and/or the sensitivity to diffuse emission at the frequency of radio 
imaging \citep[see, e.g.,][]{2007AJ....133.2097C, 2009ApJS..181..548C, 2003AJ....126..113W}. As 
an example, the XRG candidates J0941$+$2147, J1206$+$3812 and J1444$+$4147 were included in C07 list 
despite their showing abnormally short wings. However, the wings stood out clearly in subsequent 
VLA 1.4\,GHz images by \citet{2018ApJ...852...47R}. Guided by this, we have included in our XRG catalog 
several candidates (e.g., J0028$-$0026 and J0930$+$2343) whose secondary lobes are visible, albeit no 
more in size than $\sim$50\% of the primary lobes, provided a hint of inversion symmetry is discernible 
in their existing radio maps (TGSS\_ADR1 and/or FIRST). The classification `strong' has been assigned  to some XRG candidates in our 
catalog, in spite of their radio sizes not being comfortably large enough to reveal their structural details
in the existing radio maps. In adopting this somewhat less stringent approach we have been guided by the 
experience of several prominent XRGs which had appeared to be only marginally convincing  in their 
earlier (lower sensitivity/resolution) radio maps. Prominent such examples include NGC\,326 \citep{1977A&AS...29..279F}; 
3C\,315 \citep[e.g.][]{1969MNRAS.145...31M, 2007MNRAS.374.1085L} and 3C\,63 \citep{1998ApJS..119...25H}.

\begin{figure}
\centering
\includegraphics[width=0.45\textwidth]{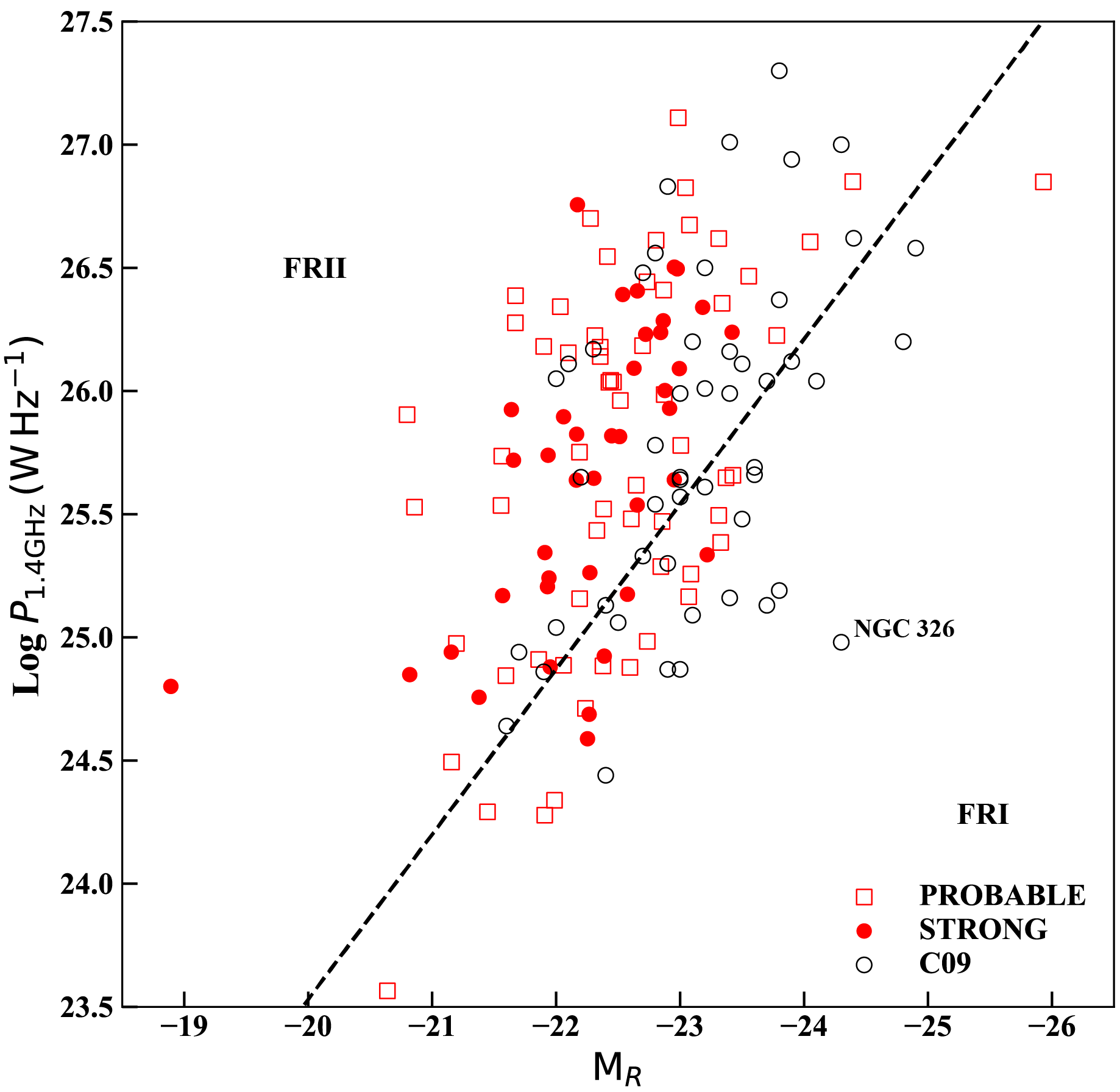}
\caption{The 1.4\,GHz radio luminosity versus absolute R-band magnitude. The probable and the strong XRG 
candidates with spectroscopic redshifts are marked with red open squares (probable) and red filled circles 
(strong). The XRG candidates recompiled in \citet{2009ApJS..181..548C} are marked with 
black open circles (C09). The Owen-Ledlow luminosity division line is from \citet{2007A&A...470..531W}. 
The representative FRI type XRG, NGC\,326 is marked.}
\label{fig:lumi}
\end{figure}

Further, it needs to be borne in mind that the observed faintness and small sizes of the wings could often 
be artifacts arising from energy losses suffered by the relativistic plasma radiating in these older/fossil
radio components. To overcome this limitation, low-frequency radio imaging with 
sufficiently high angular resolution and sensitivity are mandatory. With a typical expansion velocity of 
$\sim0.1c$ \citep[e.g.][]{1998AJ....115..960T, 2004MNRAS.351..727A, 2010A&A...510A..84M, 2012ApJ...760...77A}, 
the active lobes would take $\gtrsim10^6$\,yr to grow to their typically observed dimensions 
\citep[$\sim$30 - 60\,kpc][]{2011A&A...527A..38M, 2012A&A...544A..36M}. This duration is 
comparable to the expected radiative lifetime of the relativistic plasma in the wings at decimeter 
or shorter wavelengths, once the energy supply to the wings has been cut-off, following the putative 
spin-flip \citep{1989MNRAS.239..173G, 1994A&A...285...27K, 1997MNRAS.286..215K, 1997MNRAS.292..723K, 
2011MNRAS.413.1107M, 2016ApJ...826..132S}. The consequent fading could easily push one or both radio 
wings below the detection threshold, unless imaged at a sufficiently low radio frequency. This situation 
may not be uncommon, as exemplified by the case of J1257$+$1228 (see Figure 
\ref{fig:tgss}). All such sources could be classified as `strong' XRG candidates only because at the
low frequency (150 MHz) of the TGSS\_ADR1 images, their wings are {\it intrinsically} large enough 
to be resolved and recognized as such. In contrast, the detection of these wings in the 1.4\,GHz FIRST 
survey was hampered both due to their shorter radiative lifetimes and 
the reduced sensitivity of the FIRST survey to extended radio emission.

\subsection{The Fanaroff-Riley classification for X-shaped radio galaxies} \label{sec:fr}

\begin{figure*}
\centering
\includegraphics[width=1\textwidth]{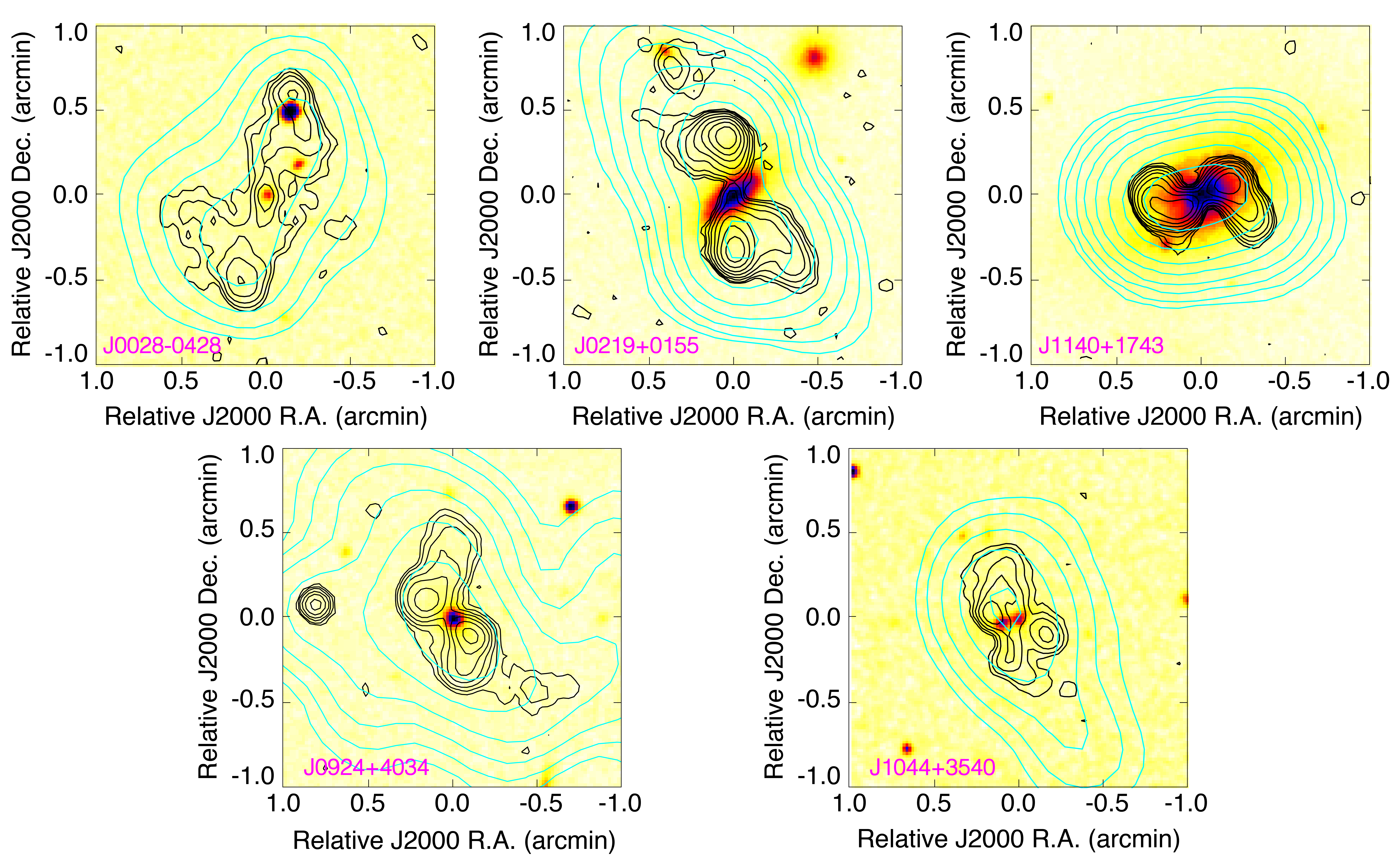}
\caption{Radio images of the five XRG candidates 
located in the FR\,I domain of the Owen-Ledlow diagram. The radio contours are overlaid on the DSS red filter images, the black contours display the FIRST 1.4 GHz maps 
and the cyan contours show the TGSS\_ADR1 images at 150\,MHz. The contours are plotted at $(1, 2, 4, 8, 16, 
32, ...)\times3\sigma$, where $1\sigma$ noise values are 0.19, 0.13, 0.11, 0.16, and 0.15 mJy\,beam$^{-1}$ for the FIRST 
maps of J0028$-$0428, J0219$+$0155, J1140$+$1743, J0924$+$4034 and J1044$+$3540, respectively, and 3.5 mJy\,beam$^{-1}$ for all the TGSS\_ADR1 maps.}
\label{fig:xrgfri}
\end{figure*}

Radio luminosities of X-shaped radio galaxies have been found to fall near the FR\,I/FR\,II 
division \citep[e.g.,][]{2002MNRAS.330..609D}. This trend is further strengthened by the present 
catalog of XRG candidates (section \ref{sec:radio}). In the model of \citet[][]
{2003ApJ...594L.103G, 2012RAA....12..127G} this property is linked to the formation mechanism of XRGs 
itself. Going a step further, \citet{2009ApJS..181..548C} have shown that the XRGs are located near the 
FR\,I/FR\,II dividing line in the `radio-optical luminosity diagram', also known as the Owen-Ledlow plane 
\citep[][]{1996AJ....112....9L}. We now briefly examine this issue in relation to the present sample of 
XRG candidates. It is well known that the active (primary) radio lobes of XRGs are mostly of the FR\,II 
type \citep{1992ersf.meet..307L, 2018ApJ...852...48S}, although some XRGs with FR\,I primary lobes do exist 
\citep{1984MNRAS.210..929L, 1992ApJS...80..137J, 2001A&A...380..102M}. Figure \ref{fig:lumi} shows 
the distribution of our XRG candidates (both `strong' and `probable' types) on the Owen-Ledlow plane.
For this we have only used a subset of 101 sources from our sample having spectroscopic redshifts and NVSS 1.4\,GHz flux density, 
and augmented the sample by including the 50 XRGs with spectroscopic redshifts, taken from the compilation of 
\citet[][]{2009ApJS..181..548C}. The FR dividing line shown in Figure \ref{fig:lumi} ican be parametrized 
as $logP_{1.4\mathrm{GHz=-0.67M_R+10.13}}$, where $P_\mathrm{1.4GHz}$ is the radio luminosity at 
1.4\,GHz and $M_R$ is the $r$-band absolute magnitude of the host galaxy \citep{2007A&A...470..531W}. 
It is evident that our XRG candidates are clustered near the dividing line defined above, in accord with C09. 
Furthermore, a vast majority of our XRG candidates ($\sim$80\% to 90\%) falls above the dividing line, in the region known to 
be populated predominantly by FR\,II sources. This is not unexpected, given that the XRG candidates (both 
`strong' and `probable') in our catalog are mostly consistent with the FR\,II morphology. Specifically, we 
find that only 5 out of the 39 `strong' XRG candidates from our sample with spectroscopic redshifts and NVSS 1.4\,GHz flux density, fall 
below the Owen-Ledlow dividing line (i.e., in the preferred domain of FR\,I sources). An inspection of 
their FIRST radio maps shows that J0956$+$0001 in fact has a clear FR\,II morphology, whereas J0727$+$3956, 
J1145$+$1529 and J1407$+$2722 are too small for a reliable morphological classification. Only the source, 
J0924$+$4034 has a morphology (see the bottom-left panel of Figure \ref{fig:xrgfri}) clearly reminiscent of the prototypical FR\,I XRG 
NGC\,326 \citep[which is marked in Figure \ref{fig:lumi}, see][]{1977A&AS...29..279F}.

We now turn to the 17 `probable' XRG candidates falling in the FR\,I region of the Owen-Ledlow diagram, 
i.e., below the dividing line in Figure \ref{fig:lumi}. Based on the available maps, we found three of them  
(J0028$-$0428, J0219$+$0155 and J1140$+$1743, see the top row of panels of Figure \ref{fig:xrgfri}) to show an (inner) morphology reminiscent 
of the FR\,I type XRG NGC\,326 \citep[see Figure 6 in][]{2001A&A...380..102M}. J1140$+$1743, also known as 
NGC\,3801, has earlier been identified as a Z-shaped source \citep{2009ASPC..407..104H}. We have classified 
it as a probable XRG (or, an intermediate between the X- and Z-shaped radio galaxies) based on its FIRST image 
shown in the top-right panel of Figure \ref{fig:xrgfri}. As it is known to be a post-merger star-forming galaxy 
\citep[][]{2012MNRAS.422L..38H}, its wings could even have formed due to pre-merger gravitational interaction 
between the two galaxies. Coming to some other `probable' XRG candidates in the FR\,I domain (Figure \ref{fig:lumi}), we
find that a few of them (e.g. J1044$+$3540, see the bottom-right panel of Figure \ref{fig:xrgfri}) appear remarkably similar in radio morphology 
to the early low resolution radio map of XRG NGC\,326 \citep{1977A&AS...29..279F}, which is a classical example of FR\,I XRG, on account of its prominent twin jets 
\citep{2001A&A...380..102M}. In relation to such sources, \citet{2002Sci...297.1310M} suggested that if the 
spin of the active SMBH undergoes a reorientation due to an impulsive torque of external origin, the resulting 
flip of its spin vector could lead to an X-shaped radio morphology \citep[see also][]
{1978MNRAS.185..527B, 2001A&A...380..102M}. On the other hand, a S- or Z-shaped radio morphology would result
if the torque operates on the SMBH only gradually, i.e., its duration is longer than the jet outflow time scale. 
In the case of the radio galaxy NGC\,326, an impulsive torque is, in fact, quite plausible, since its 
host is a dumb-bell galaxy comprised of a pair of massive ellipticals in gravitational interaction 
\citep{1980A&A....85..101B}. In this scenario, one may expect to find sources with an intermediate morphology 
(between S/Z-shape and X-shaped), in case the outflow and spin reorientation time scales are comparable.

\begin{figure*}
\centering
\includegraphics[width=1\textwidth]{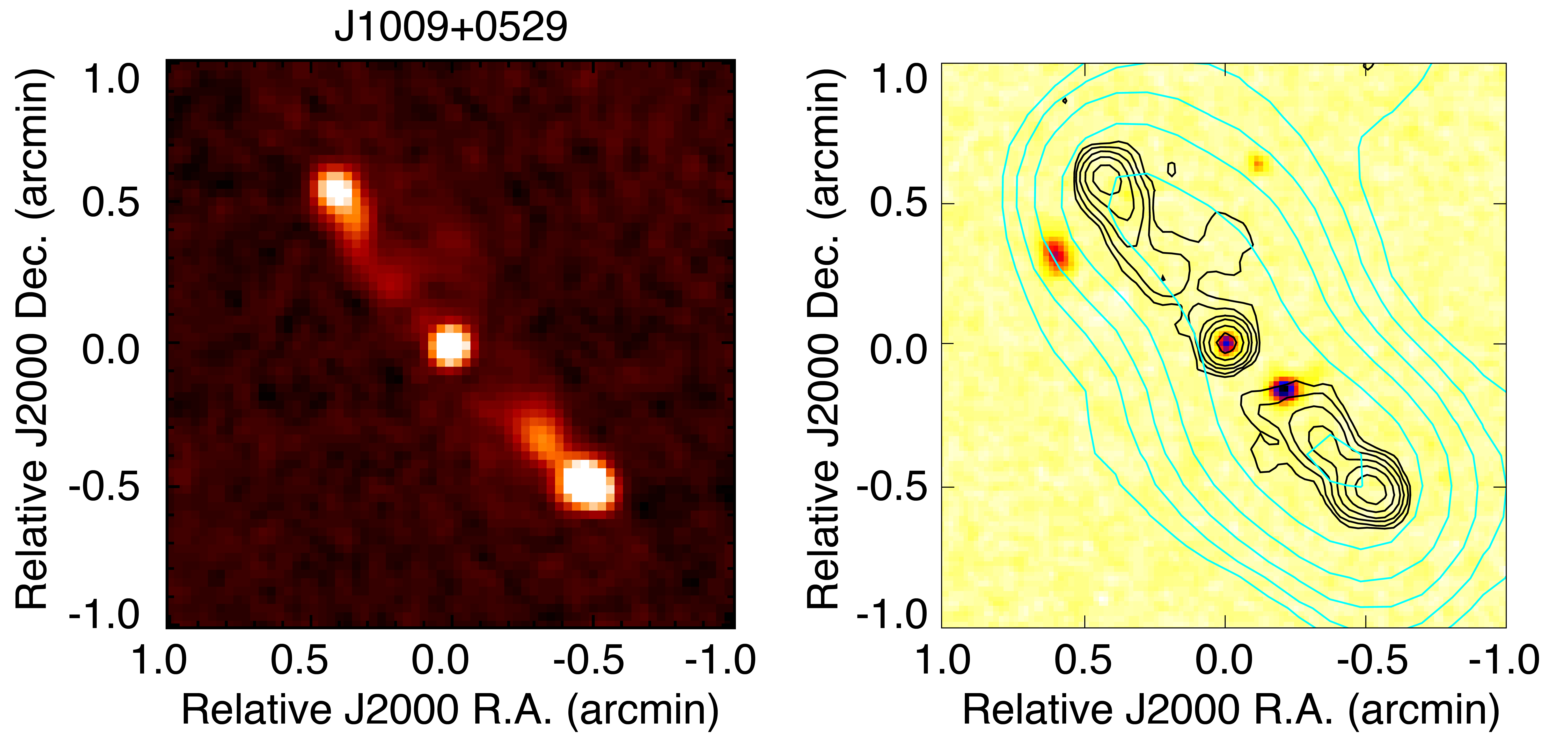}
\caption{The `probable' XRG candidate J1009$+$0529 in our catalog, the fields are 
centered on the optical counterparts. The left panel shows the FIRST grayscale image and the right panel shows the FIRST (black) and TGSS\_ADR1 (cyan) contours 
overlaid on the DSS red filter image. Contours levels are set at (1, 2, 4, 8, 16, 32, ...)$\times3\sigma$, where $1\sigma$ rms noise is 0.18\,mJy\,
beam$^{-1}$ for the FIRST map and 3.5\,mJy\,beam$^{-1}$ for the TGSS\_ADR1 map.}
\label{fig:1009+0529}
\end{figure*}

Speculating on the later evolution of the reoriented (Z/S- and X-shaped) radio sources, 
\citet{2002Sci...297.1310M} proposed that such sources would eventually evolve into FR\,II type, on a time scale 
of $\sim10^8$ years. For instance, in the FIRST image of J1009$+$0529 (Figure \ref{fig:1009+0529}), 
the large brightness contrast seen between the active radio lobes and the (much older) wings might be indicative of such a transition in
making, although this remains to be confirmed through spectral index imaging. On the other hand, \citet{2009ApJ...695..156S} have proposed an opposite evolutionary scheme (from FR\,II to FR\,I), motivated by the examples of FR\,I radio galaxies in which a newly formed inner lobe pair has been detected, such as 3C\,315 and B2014$-$558. In these sources, whereas the extended primary lobes are of FR\,I type \citep{1984MNRAS.210..929L,1992ApJS...80..137J}, 
their inner doubles show an FR\,II morphology \citep{2008arXiv0806.3518S, 2000ApJS..129...33D}, see also MERLIN 
1.6\,GHz images of 3C\,315 (Sanghera \& Leahy, unpublished, see \textcolor{blue}{http://www.jb.man.ac.uk/atlas/other/3C315.html}). However, while the putative evolutionary transition from FR\,II to FR\,I morphology may indeed be physically plausible \citep[see also][]{1988Natur.333...49G}, it becomes imperative when the few observed FR\,I type XRGs are sought to 
be explained in terms of the backflow deflection model \citep[see][]{2009ApJ...695..156S, 2012ApJS..199...27S}.

\section{Summary and future work} \label{sec:sum}

Based on a careful visual inspection of the radio sources in the latest release of the VLA FIRST Survey at 1.4 GHz, we have presented here a catalog of 290 winged or X-shaped radio galaxies (XRGs), which almost triples the number of XRG candidates cataloged by \citet{2007AJ....133.2097C}. We classify 106 of them as `strong' and 184 as `probable' XRG candidates, which reflects that roughly $\sim$2 to 6\% of the radio sources are XRGs. The strong candidates are most likely to be confirmed by follow-up radio observations, while the probable candidates would also be particularly useful for designing future search campaigns for XRGs. The present work extends to smaller angular sizes the XRG search conducted by \citep{2007AJ....133.2097C}, which revealed 100 XRG candidates with a high confirmation rate 
through follow-up observations. The radio structural information on our XRG candidates has been taken mainly from  
the FIRST (1.4\,GHz), NVSS (1.4\,GHz) and TGSS\_ADR1 (150\,MHz) surveys. Combining this information 
with the SDSS data, we were able to find optical counterparts for 85 of our 106 strong XRG candidates and 145 of 
our 184 probable XRG candidates. The corresponding numbers of sources with spectroscopic redshifts are 41 
(48\%) and 61 (42\%), respectively, and the median redshifts are 0.37 and 0.41, respectively. Nine quasars are found among the 106 strong XRG candidates. The sample also contains two strong XRGs showing double-peaked optical emission lines ($z =$ 0.19 and 0.33), which makes them excellent candidates for AGN harboring binary supermassive black holes. 

On the radio side, as expected, a vast majority of the present strong XRG candidates are 
found to have steep radio spectra between 150\,MHz and 1.4\,GHz, and also between 1.4\,GHz and 5\,GHz (median $\alpha\sim-0.81$). However, a rather flat radio spectrum ($\alpha > -0.3$) has been found for 10 of our strong XRG candidates. At present, none of them are known to be a quasars. 
Recently, \citet{2019arXiv190311632L} have shown that in terms of radio spectral index, active lobes do not seem to differ from the (fainter) wings, which may likely support the twin AGN model where the spectra of the primary lobes
and secondary wings are expected to be uncorrelated.  It will be interesting to carry out a similar campaign for our bona-fide XRGs, in order to place tighter
constraints on the models of X-shaped radio sources.

To the extent permitted by the partially complete redshift and radio structural information, we find that the 
present enlarged sample of XRG candidates adheres to the previously known trend according to which XRGs cluster
around the radio luminosity dividing the FR\,I and FR\,II radio sources. More specifically, we find that out of
our 40 strong XRG candidates with known spectroscopic redshifts and NVSS 1.4\,GHz flux, as many as 35 lie above the FR division in the
Owen-Ledlow diagram. Even for the 5 objects falling below the dividing line, an FR\,II morphology seems quite plausible
in most cases. Together with the previously reported samples of XRGs \citep{2007AJ....133.2097C, 2011ApJS..194...31P}, the present large sample would allow a closer look into the relative occurrence of X-shaped and Z-shaped radio morphologies among extragalactic radio sources and, in particular, the occurrence of Z-symmetry among the radio wings in XRGs \citep{2003ApJ...594L.103G}. In parallel, we are studying possible relationship of the radio structure to the properties of the host galaxy (Joshi et al., in prep.).


\acknowledgments

This work was supported by the National Key R\&D Program of China (2016YFA0400703, 2018YFA0404603). 
LCH is supported by National Key R\&D Program of China (2016YFA0400702) and the National Science 
Foundation of China (11721303). XFY is supported by Xinjiang Tianchi Bairen project. 
This work makes use of public data from NSF's Karl G. Jansky Very Large Array (VLA), the VLA facility 
is operated by National Radio Astronomy Observatory (NRAO). The National Radio Astronomy Observatory 
is a facility of the National Science Foundation operated under cooperative agreement
by Associated Universities, Inc. This work uses the public data from GMRT TGSS survey, we thank the staff of 
the GMRT that made these observations possible. GMRT is run by the National Centre for Radio Astrophysics 
of the Tata Institute of Fundamental Research. This work makes use of the Solan Digital Sky Survey data, funding for the 
Sloan Digital Sky Survey (SDSS) has been provided by the Alfred P. Sloan Foundation, the 
Participating Institutions, the National Aeronautics and Space Administration, the National Science Foundation, 
the U.S. Department of Energy, the Japanese Monbukagakusho, and the Max Planck Society. The SDSS Web 
site is http://www.sdss.org/. The SDSS is managed by the Astrophysical Research Consortium (ARC) for the 
Participating Institutions. The Participating Institutions are The University of Chicago, Fermilab, the Institute for 
Advanced Study, the Japan Participation Group, The Johns Hopkins University, Los Alamos National Laboratory, 
the Max-Planck-Institute for Astronomy (MPIA), the Max-Planck-Institute for Astrophysics (MPA), New Mexico State 
University, University of Pittsburgh, Princeton University, the United States Naval Observatory, and the University 
of Washington.

%






\bibliographystyle{aasjournal}
\bibliography{XRG_11May_v3}

\appendix


\begin{longtable}{lccclcccccccrr}
\renewcommand\arraystretch{1}

\label{tab:strong} \\
\caption{The `strong' X-shaped radio source candidates}   \\
\toprule                                                                                                                                                                                   
\multicolumn{1}{c}{Short Name}   &   R.A.     &    Dec.    &    z     &   \multicolumn{1}{c}{Qual.}    & $r$  &  $M_R$     &   $S_{i,0.15}$  &  $S_{i,1.4}$  &   $S_{i,5}$  &    Ref.$_{5\mathrm{GHz}}$     & \multicolumn{1}{c}{$\alpha_{0.15}^{1.4}$}    & \multicolumn{1}{c}{$\alpha_{1.4}^{5}$}           \\
                                                     &      (J2000)      &      (J2000)       &         &            &             &     &   ($\mathrm{Jy}$)     &      ($\mathrm{Jy}$)         &  ($\mathrm{Jy}$)          &    & &                                             \\
\multicolumn{1}{c}{(1)}                  &   (2)      &    (3)     &   (4)    &   \multicolumn{1}{c}{(5)}      & (6)  &  (7)      &  (8) &   (9)            &  (10)         &   (11)        & \multicolumn{1}{c}{(12)} &  \multicolumn{1}{c}{(13)}                                 \\
\hline                                                                                                                                                    
\endfirsthead

\caption{Continued}   \\
\toprule                                                                                                                                                                                   
\multicolumn{1}{c}{Short Name}   &   R.A.     &    Dec.    &    z     &   \multicolumn{1}{c}{Qual.}    & $r$  &  $M_R$     &   $S_{i,0.15}$  &  $S_{i,1.4}$  &   $S_{i,5}$  &    Ref.$_{5\mathrm{GHz}}$     & \multicolumn{1}{c}{$\alpha_{0.15}^{1.4}$}    & \multicolumn{1}{c}{$\alpha_{1.4}^{5}$}               \\
               &      (J2000)     &       (J2000)       &         &            &             &     &  ($\mathrm{Jy}$)      &      ($\mathrm{Jy}$)         &  ($\mathrm{Jy}$)          &     &&                                             \\
\multicolumn{1}{c}{(1)}          &   (2)      &    (3)     &   (4)    &   \multicolumn{1}{c}{(5)}      & (6)  &  (7)      &  (8) &   (9)            &  (10)         &   (11)        & \multicolumn{1}{c}{(12)} &  \multicolumn{1}{c}{(13)}                                    \\
\hline                                                                                                                              
\endhead

\hline
\endfoot

\hline
\endlastfoot

J0004$+$1248        &$ 00\,04\,50.27  $&$ +12\,48\,39.5  $&$     0.143   $&   SPEC   &$  17.44 $&$  -21.37  $&$   ...     $&$   0.107   $&$     0.407   $&     NRAO   &$ ...    $&$  1.07  $  \\                                                                                              
J0028$-$0026$^*$    &$ 00\,28\,28.94  $&$ -00\,26\,24.6  $&$     0.287   $&   PHOT   &$  18.96 $&$  -21.52  $&$   ...     $&$   ...     $&$     0.390   $&     NRAO   &$ ...    $&$ ...    $  \\ 
J0030$+$1121        &$ 00\,30\,23.86  $&$ +11\,21\,12.5  $&$     0.449   $&   SPEC   &$  20.16 $&$  -22.06  $&$   0.515   $&$   0.114   $&$     0.136   $&     NRAO   &$ -0.67  $&$  0.14  $  \\ 
J0121$+$0051        &$ 01\,21\,01.23  $&$ +00\,51\,00.3  $&$     0.238   $&   SPEC   &$  17.80 $&$  -22.16  $&$   1.124   $&$   0.267   $&$     0.105   $&     NRAO   &$ -0.64  $&$ -0.75  $  \\ 
J0216$+$0244        &$ 02\,16\,35.79  $&$ +02\,44\,00.9  $&$     0.184   $&   PHOT   &$  17.58 $&$  -21.70  $&$   2.650   $&$   0.300   $&$     0.248   $&     PMN    &$ -0.97  $&$ -0.15  $  \\ 
J0319$-$0202$^*$    &$ 03\,19\,37.58  $&$ -02\,02\,48.7  $&$     ...     $&   ...    &$  ...   $&$  ...     $&$   0.849   $&$   0.212   $&$     0.080   $&     NRAO   &$ -0.62  $&$ -0.78  $  \\ 
J0710$+$3546        &$ 07\,10\,31.14  $&$ +35\,46\,49.8  $&$     ...     $&   ...    &$  ...   $&$  ...     $&$   ...     $&$   0.073   $&$     0.030   $&     NRAO   &$ ...    $&$ -0.71  $  \\ 
J0715$+$4910        &$ 07\,15\,10.12  $&$ +49\,10\,53.2  $&$     ...     $&   ...    &$  ...   $&$  ...     $&$   0.513   $&$   0.084   $&$     0.044   $&     NRAO   &$ -0.81  $&$ -0.52  $  \\
\end{longtable}

\begin{flushleft}                                                                                                                    
                                                                                                                                     
\textbf{Notes}: \\                                                                                                                   
(1) short name; (2) Right Ascension (J2000, hh:mm:ss); (3) Declination (J2000, dd:mm:ss); (4) redshift;\\                            
(5) redshift quality flag. PHOT: photometric redshift; SPEC: spectroscopic redshift; SPEC/Q: quasar; \\                                              
(6) apparent $r$-band magnitude of the host galaxy; \\                                                                                      
(7) absolute $R$-band magnitude of the host galaxy; \\                                                                                      
(8) integrated $\mathrm{150\,MHz}$ flux density (Jy, the uncertainty is 10\% of the total flux density); \\                              
(9) integrated $\mathrm{1.4\,GHz}$ flux density (Jy, the uncertainty is 3\% of the total flux density); \\                              
(10) integrated $\mathrm{5\,GHz}$/$\mathrm{4.85\,GHz}$ flux density (Jy); \\                                                         
(11) references for $\mathrm{5\,GHz}$/$\mathrm{4.85\,GHz}$ flux density, NRAO: NRAO Green Bank 300 Foot $\mathrm{4.85\,GHz}$ survey; 
PMN: The Parkes-MIT-NRAO 4.85\,GHz Surveys; MIT-GB: The MIT-Green Bank $\mathrm{5\,GHz}$ Survey. \\  
(12) spectral index between $\mathrm{150\,MHz}$ and $\mathrm{1.4\,GHz}$  \\
(13) spectral index between $\mathrm{1.4\,GHz}$ and $\mathrm{5\,GHz}$/$\mathrm{4.85\,GHz}$ \\
$^*$ also appeared in \citet{2011ApJS..194...31P}.

\end{flushleft}   

\begin{longtable}{lccclccccccc}
\renewcommand\arraystretch{1}

\label{tab:prob} \\
\caption{The `probable' X-shaped radio source candidates}   \\
\toprule                                                                                                                                                                                   
 \multicolumn{1}{c}{Short Name}   &   R.A.     &    Dec.    &    z     &    \multicolumn{1}{c}{Qual.}    & $r$  &  $M_R$     &   $S_{i,0.15}$  &  $S_{i,1.4}$  &   $S_{i,5}$  &    Ref.$_{5\mathrm{GHz}}$                    \\
               &      (J2000)     &       (J2000)       &         &            &             &     &    ($\mathrm{Jy}$)    &      ($\mathrm{Jy}$)         &  ($\mathrm{Jy}$)          &                                                   \\
 \multicolumn{1}{c}{(1)}          &   (2)      &    (3)     &   (4)    &    \multicolumn{1}{c}{(5)}      & (6)  &  (7)      &  (8) &   (9)            &  (10)         &   (11)                                            \\
\hline                                                                                                                                                    
\endfirsthead

\caption{Continued}   \\
\toprule                                                                                                                                                                                   
 \multicolumn{1}{c}{Short Name}   &   R.A.     &    Dec.    &    z     &    \multicolumn{1}{c}{Qual.}    & $r$  &  $M_R$     &   $S_{i,0.15}$  &  $S_{i,1.4}$  &   $S_{i,5}$  &    Ref.$_{5\mathrm{GHz}}$                    \\
               &      (J2000)    &        (J2000)       &         &            &             &     &   ($\mathrm{Jy}$)     &      ($\mathrm{Jy}$)         &  ($\mathrm{Jy}$)          &                                                   \\
 \multicolumn{1}{c}{(1)}          &   (2)      &    (3)     &   (4)    &    \multicolumn{1}{c}{(5)}      & (6)  &  (7)      &  (8) &   (9)            &  (10)         &   (11)                                            \\
\hline                        
\endhead                      
                              
\hline                        
\endfoot                      
                              
\hline                        
\endlastfoot

J0009$+$0457        &$    00\,09\,18.92 $&$ +04\,57\,53.2  $&$   0.545  $&  PHOT     &$   21.20  $&$ -21.55   $&$  1.402   $&$   0.320    $&$    0.094   $&     NRAO          \\                                                                                      
J0028$-$0428        &$    00\,28\,15.26 $&$ -04\,28\,55.3  $&$   0.267  $&  SPEC     &$   17.51  $&$ -22.84   $&$  0.484   $&$   0.092    $&$    0.063   $&     PMN           \\
J0031$-$0421        &$    00\,31\,41.91 $&$ -04\,21\,56.0  $&$   0.485  $&  SPEC     &$   21.05  $&$ -21.56   $&$  0.133   $&$   0.066    $&$    ...     $&     ...           \\
J0049$-$0507        &$    00\,49\,55.56 $&$ -05\,07\,04.3  $&$   ...    $&  ...      &$   22.40  $&$ ...      $&$  0.596   $&$   0.152    $&$    0.064   $&     PMN           \\
J0051$+$0728        &$    00\,51\,27.90 $&$ +07\,28\,56.5  $&$   0.512  $&  PHOT     &$   20.13  $&$ -22.71   $&$  4.058   $&$   0.506    $&$    0.153   $&     NRAO          \\
J0055$-$0829        &$    00\,55\,51.49 $&$ -08\,29\,53.8  $&$   0.705  $&  SPEC     &$   21.11  $&$ -23.04   $&$  1.495   $&$   0.337    $&$    0.112   $&     PMN           \\
J0125$+$0617        &$    01\,25\,25.82 $&$ +06\,17\,31.5  $&$   0.274  $&  PHOT     &$   18.81  $&$ -21.57   $&$  1.875   $&$   0.423    $&$    0.161   $&     NRAO          \\
J0131$+$0314        &$    01\,31\,05.68 $&$ +03\,14\,52.6  $&$   0.468  $&  PHOT     &$   20.48  $&$ -21.68   $&$  0.846   $&$   0.182    $&$    0.059   $&     NRAO          \\
\end{longtable}

\begin{flushleft}                                                                                                                            
                                                                                                                                             
\textbf{Notes}: 
(1) short name; (2) Right Ascension (J2000, hh:mm:ss); (3) Declination (J2000, dd:mm:ss); (4) redshift;\\                            
(5) redshift quality flag. PHOT: photometric redshift; SPEC: spectroscopic redshift; SPEC/Q: quasar; \\                                              
(6) apparent $r$-band magnitude of the host galaxy; \\                                                                                      
(7) absolute $R$-band magnitude of the host galaxy; \\                                                                                      
(8) integrated $\mathrm{150\,MHz}$ flux density (Jy, the uncertainty is 10\% of the total flux density); \\                              
(9) integrated $\mathrm{1.4\,GHz}$ flux density (Jy, the uncertainty is 3\% of the total flux density); \\                              
(10) integrated $\mathrm{5\,GHz}$/$\mathrm{4.85\,GHz}$ flux density (Jy); \\                                                         
(11) references for $\mathrm{5\,GHz}$/$\mathrm{4.85\,GHz}$ flux density, NRAO: NRAO Green Bank 300 Foot $\mathrm{4.85\,GHz}$ survey; 
PMN: The Parkes-MIT-NRAO 4.85\,GHz Surveys; MIT-GB: The MIT-Green Bank $\mathrm{5\,GHz}$ Survey. \\  
$^*$ also appeared in \citet{2011ApJS..194...31P}.                                                                                                                                              
\end{flushleft}                                                                                                                              
\begin{figure*} 
\centering 
                                                      \includegraphics[scale=0.15]{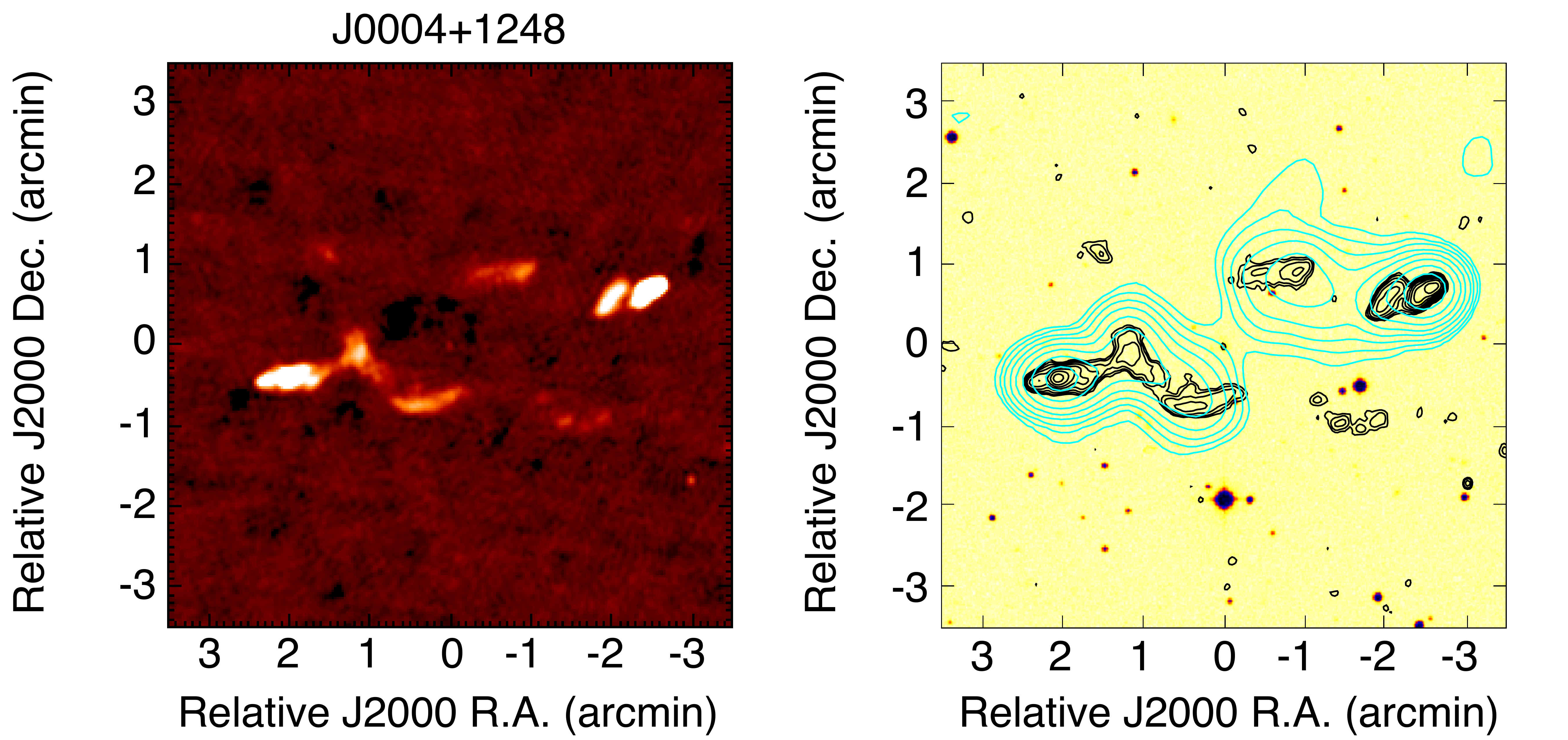}
                                                      \includegraphics[scale=0.15]{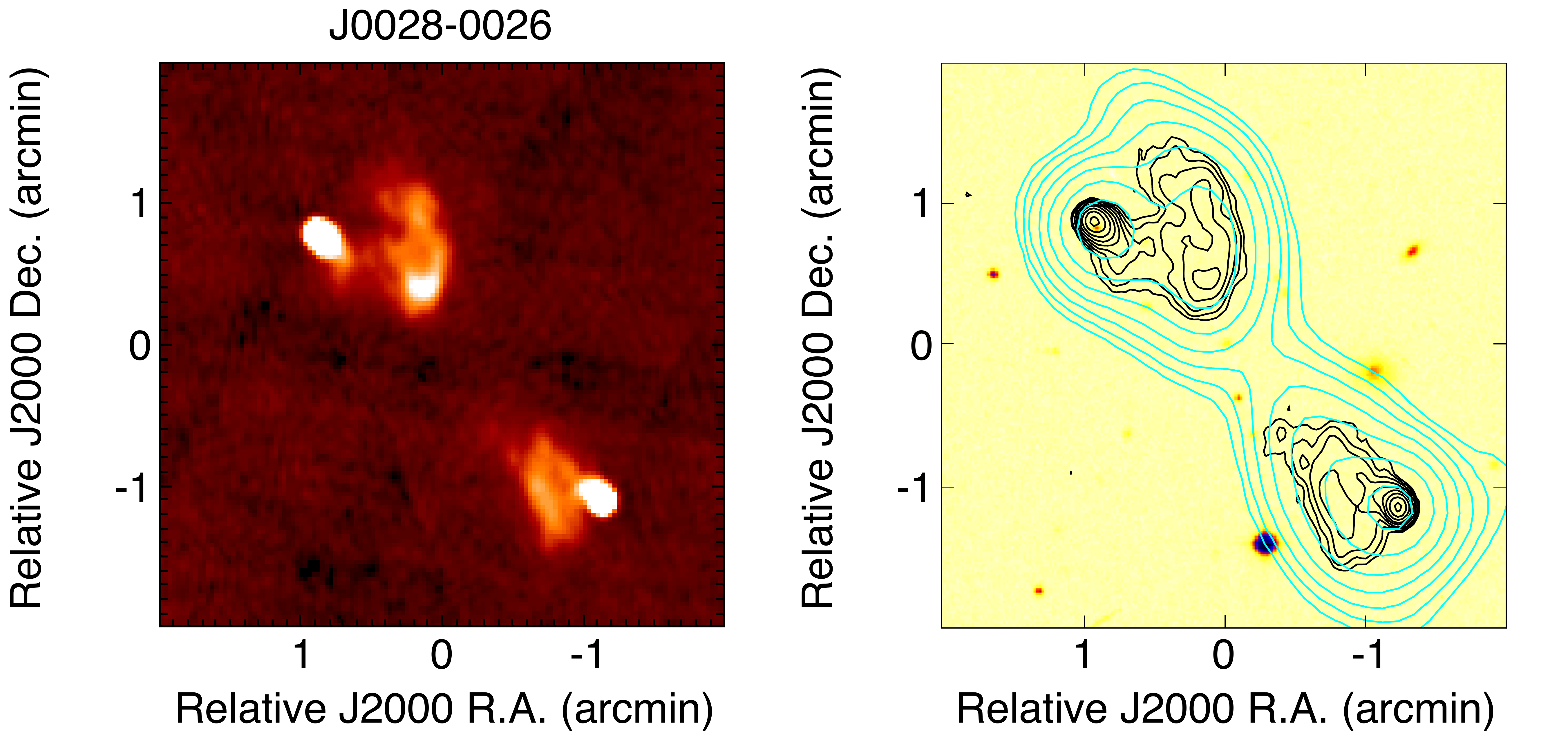}
                                                      \includegraphics[scale=0.15]{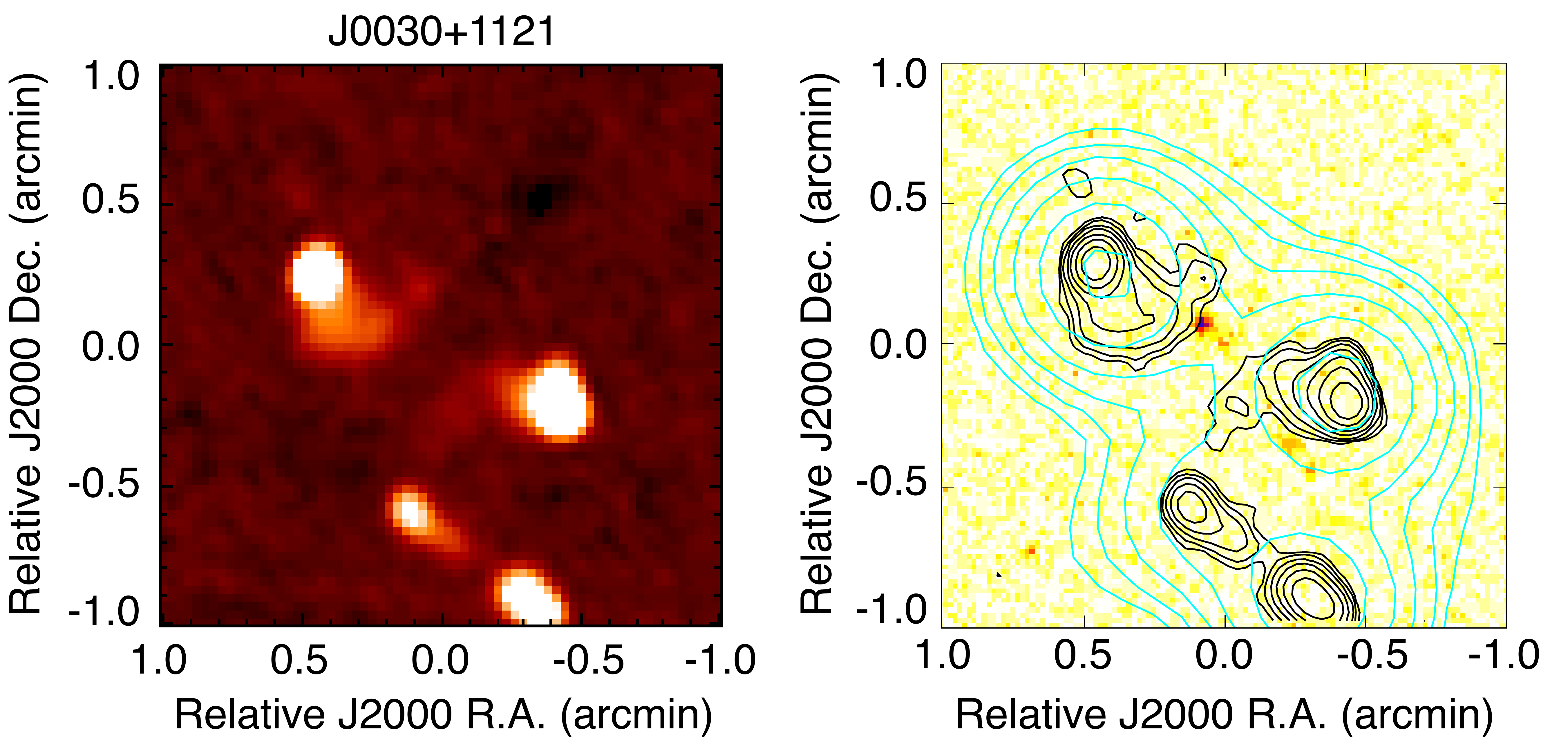}
                                                      \includegraphics[scale=0.15]{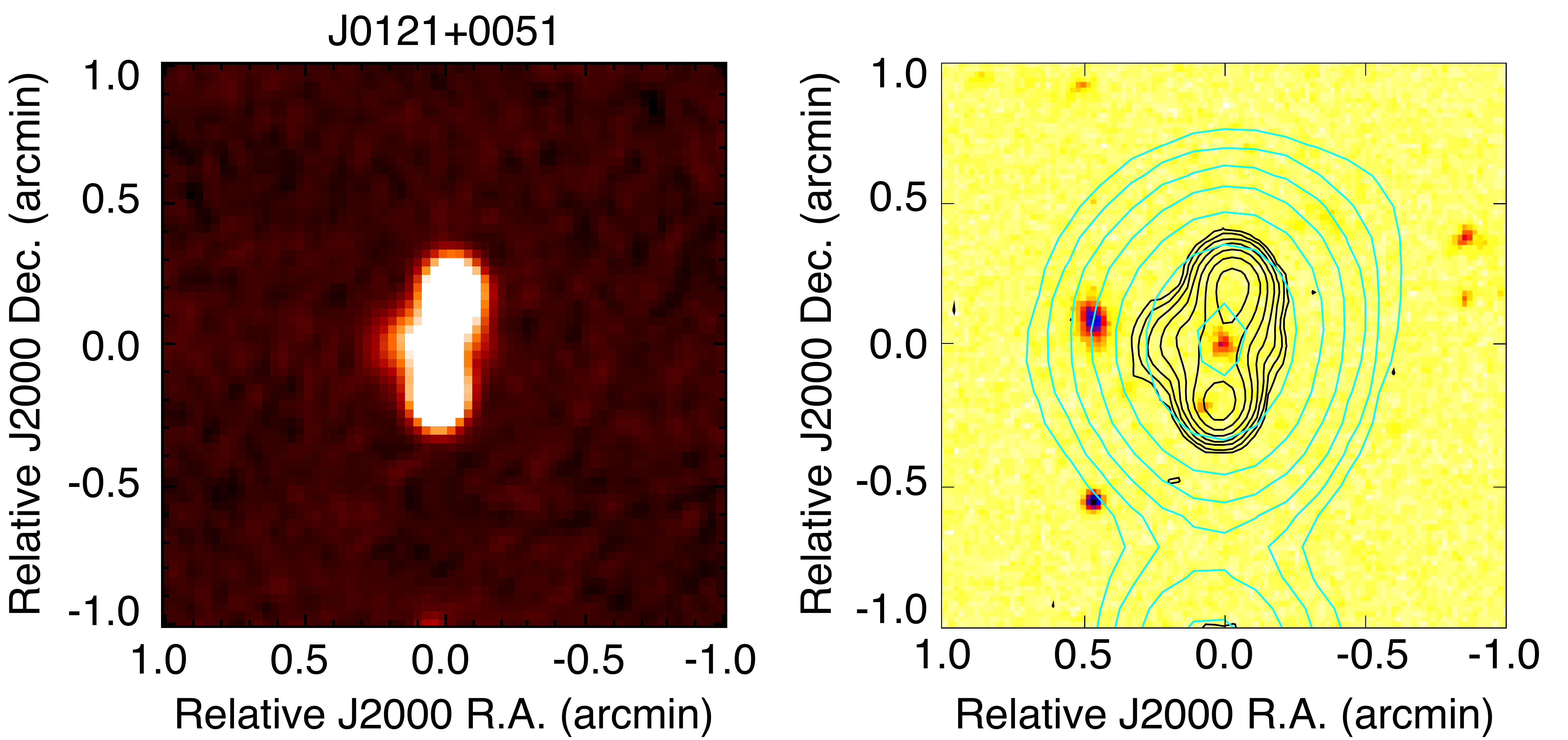}
                                                      \includegraphics[scale=0.15]{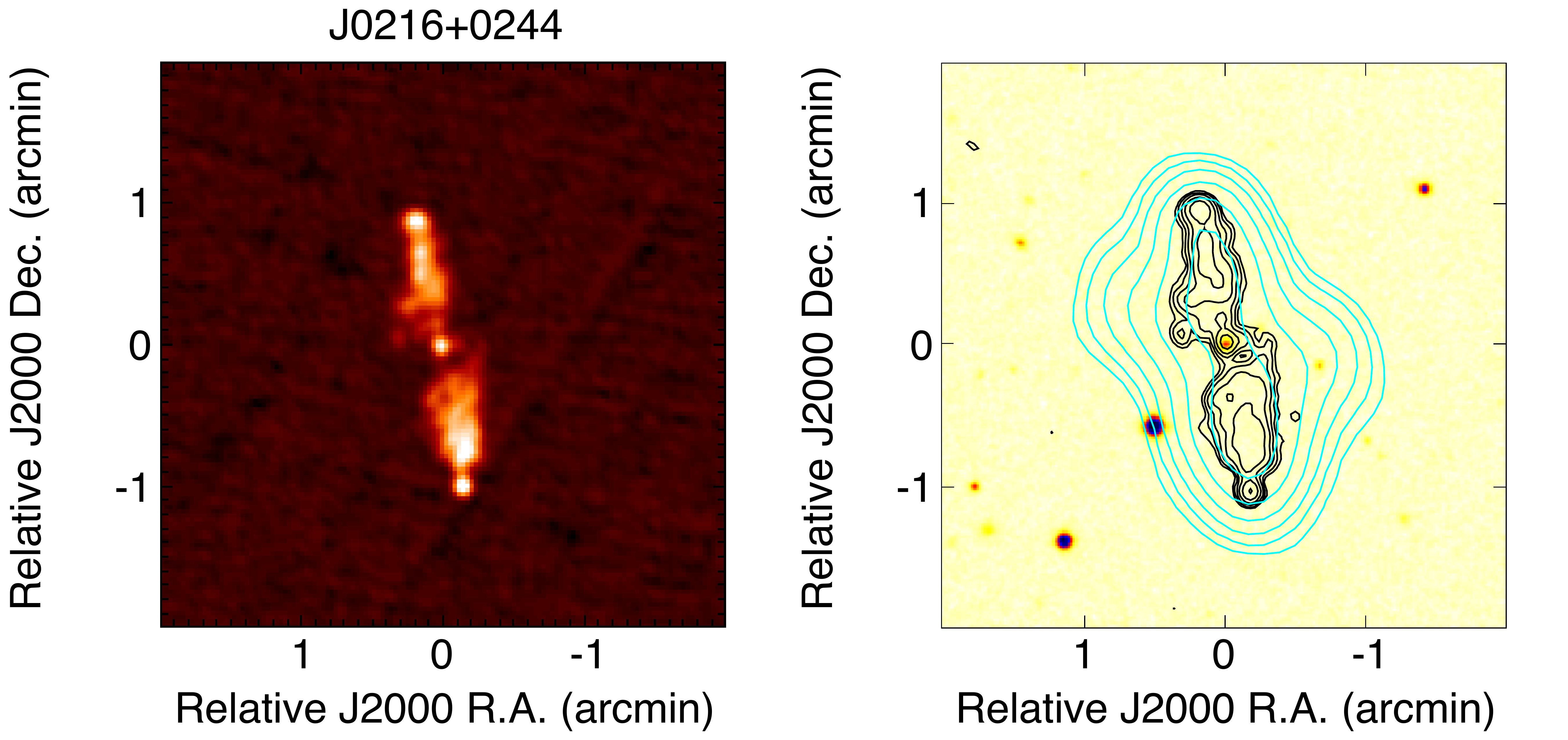}
                                                      \includegraphics[scale=0.15]{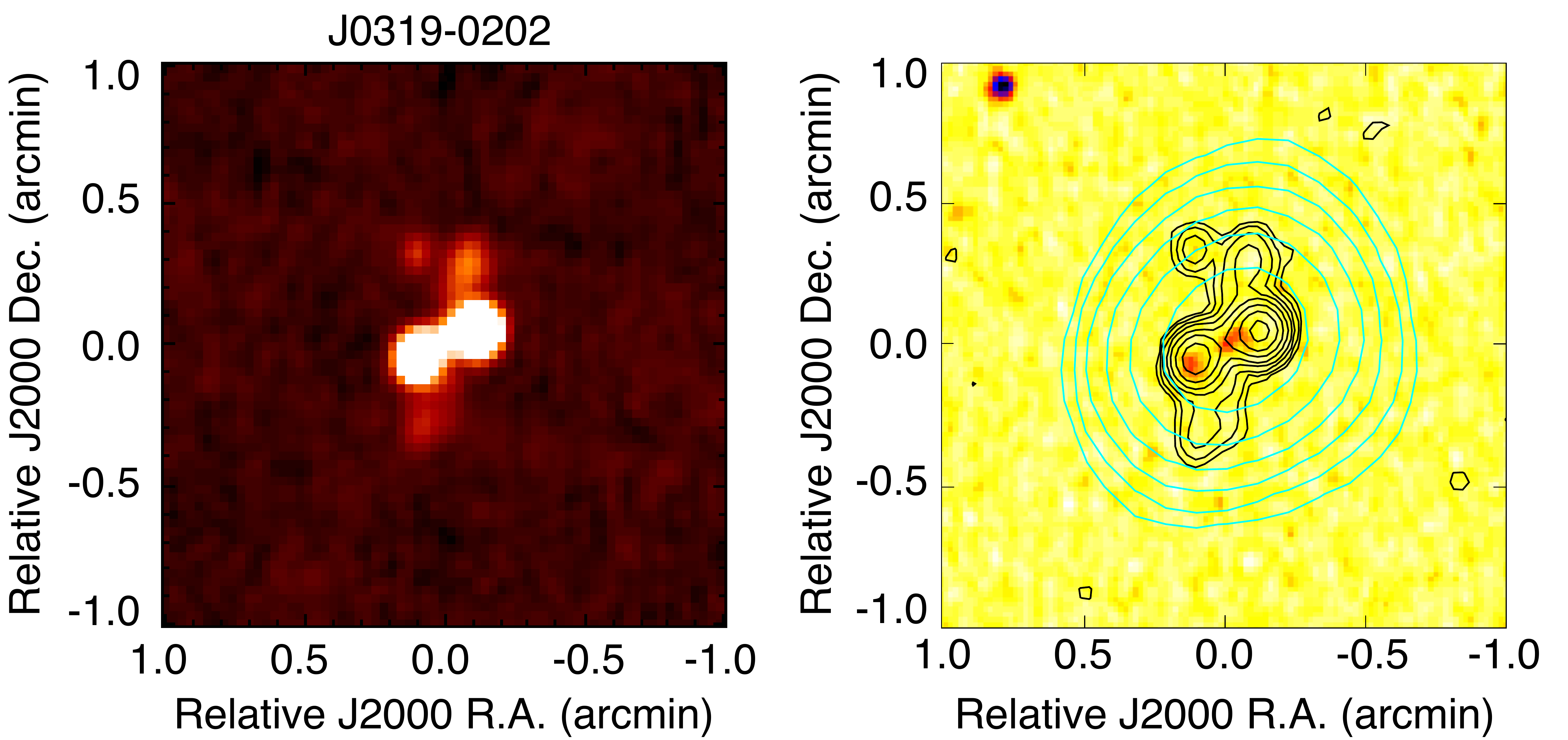}
                                                      \includegraphics[scale=0.15]{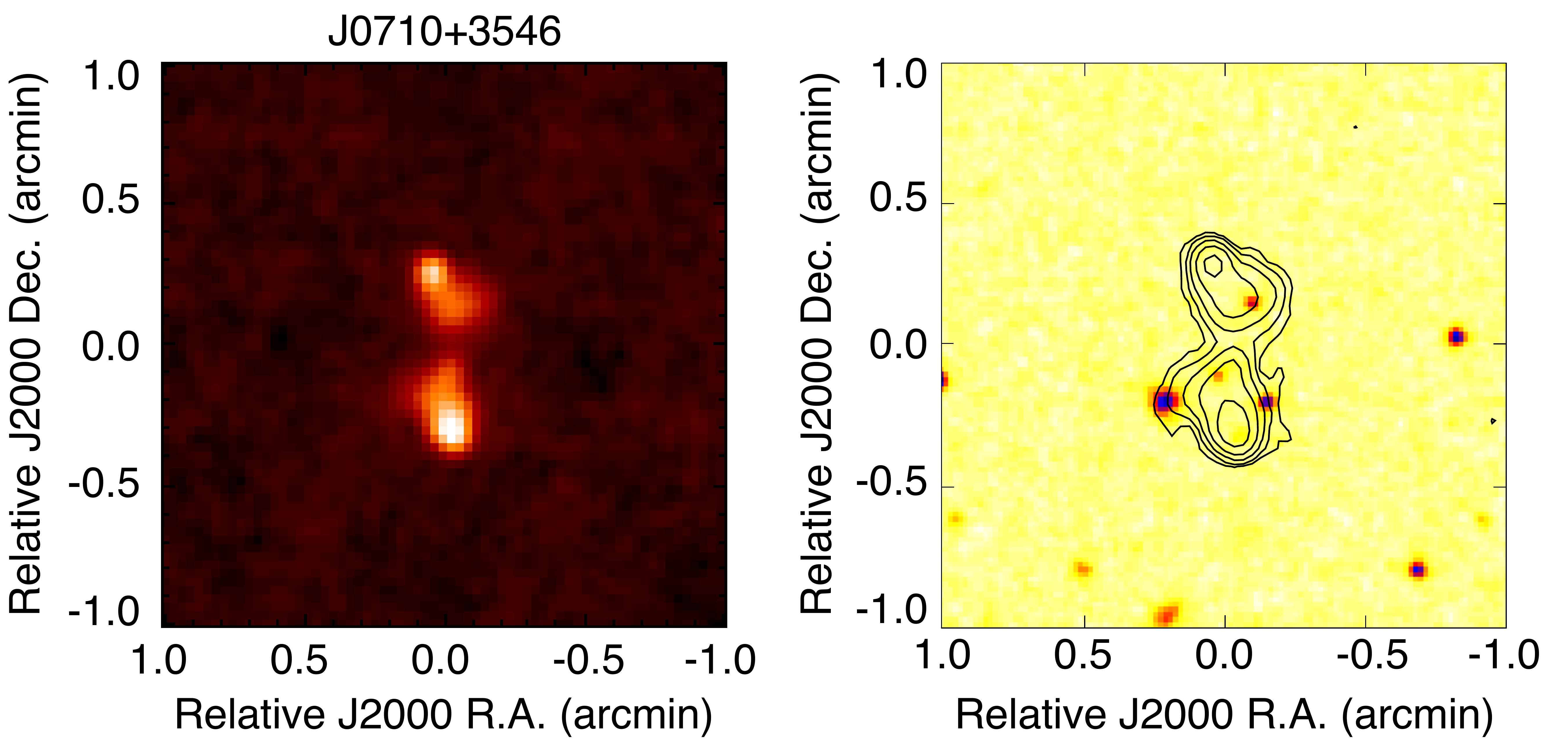}
                                                      \includegraphics[scale=0.15]{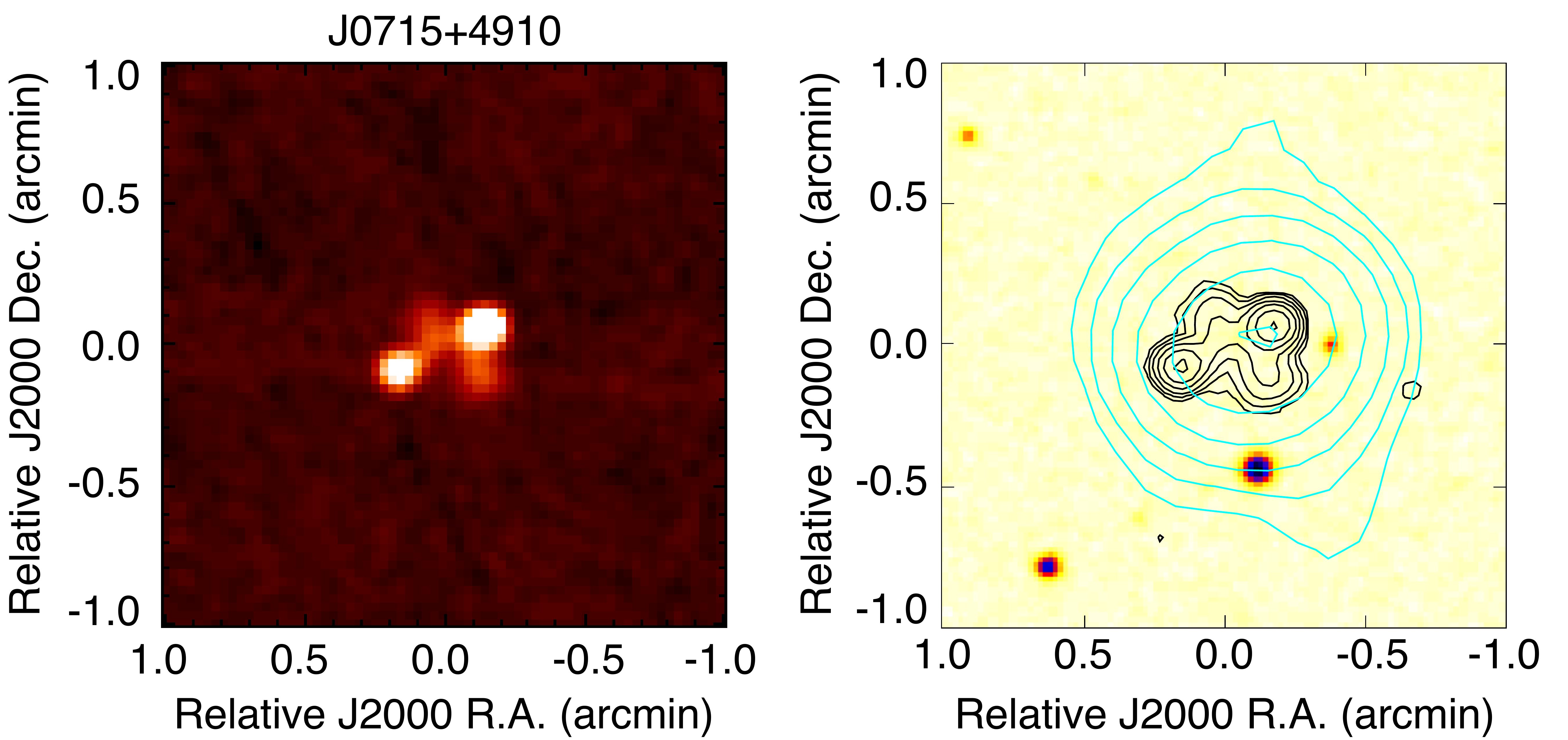}
                                                      \includegraphics[scale=0.15]{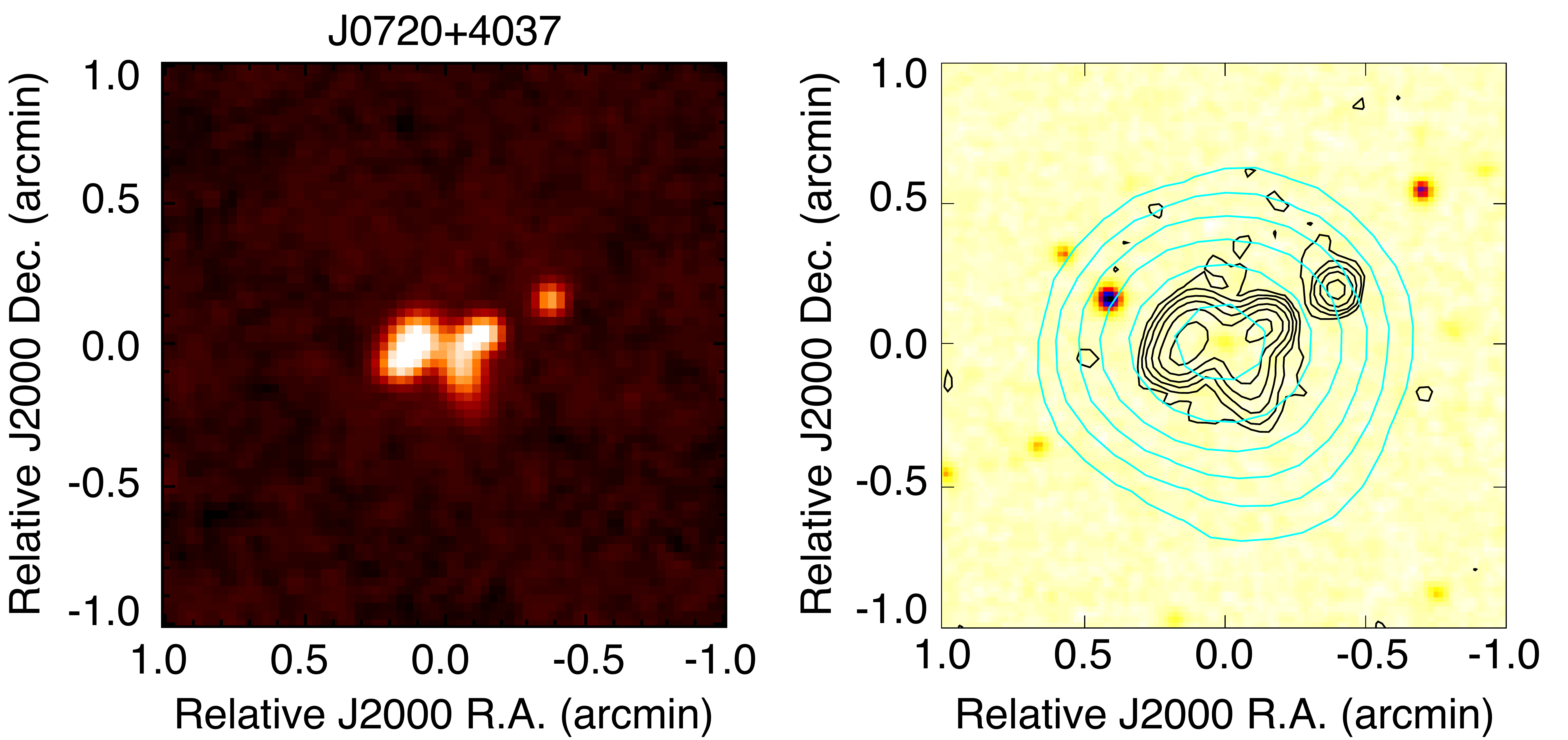}
                                                      \includegraphics[scale=0.15]{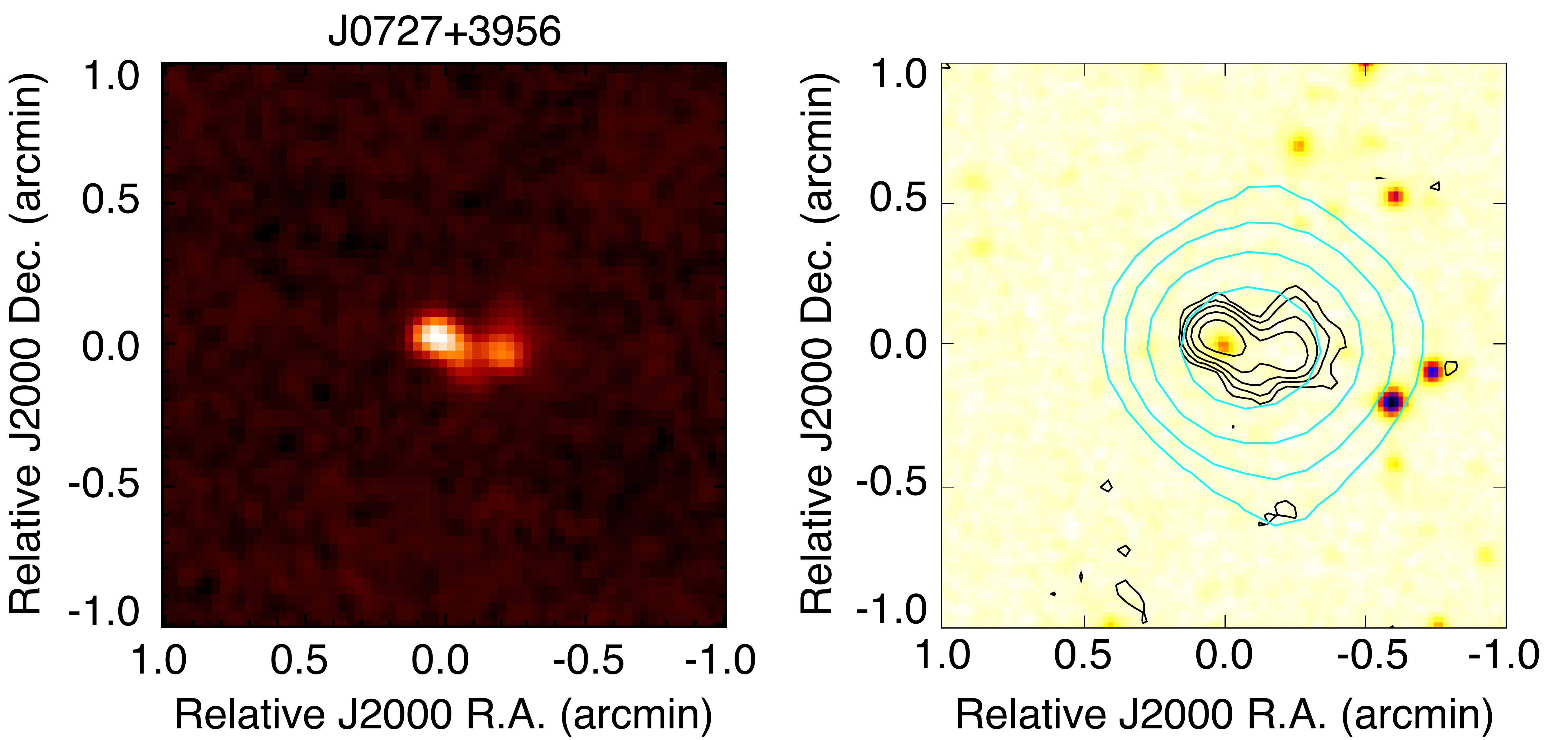}
\caption{Images of our 106 `strong' X-shaped radio galaxy candidates: VLA FIRST 1.4\,GHz images (left figure of each sub-panel) and DSS red filter images (right figure of each sub-panel) overlaid with TGSS\_ADR1 150\,MHz (cyan) and VLA FIRST 1.4GHz contours (black). The fields are centered on the optical counterparts when identified, and otherwise on positions based on the radio morphologies. The radio contours are plotted as $3\sigma\times$(1, 2, 4, 8, 16, 32, 64, ...), where the $\sigma$ is rms noise. The rms noise for the 1.4 GHz FIRST images is taken from VLA FIRST archive. In addition,  a median rms noise of 3.5\,mJy\,beam$^{-1}$ is opted for TGSS\_ADR1 fields. The images are published in its entirety in the electronic edition of the Astrophysical Journal Supplement Series.}
\label{fig:xrgs}
\end{figure*}

\end{document}